\documentclass[preprint,12pt, 3p]{elsarticle}

\usepackage{amsmath}
\usepackage{amssymb}
\usepackage{mathrsfs}
\usepackage{hyperref}
\usepackage{bm}
\usepackage{accents}
\usepackage{xcolor}
\usepackage{graphicx}
\usepackage{wrapfig}
\usepackage[table]{xcolor}
\usepackage{color,soul}

\usepackage{float}
\usepackage{placeins}
\usepackage{algorithm}
\usepackage{algpseudocode}
\usepackage{tcolorbox}
\usepackage{algorithm}
\usepackage{algpseudocode}
\tcbuselibrary{breakable}

\newtcolorbox{codeboxraw}{%
  breakable,
  colback=gray!10,
  colframe=gray!10,
  boxrule=0pt,
  left=6pt,
  right=6pt,
  top=6pt,
  bottom=6pt
}

\usepackage{caption}

\newdefinition{rmk}{Remark}

\begin{document}

\begin{frontmatter}

\title{Influence of Heterogeneity on the Response of Architected Metamaterials}

\author[1]{Sarvesh Joshi}
\author[2]{Jingye Tan}
\author[3]{Craig M. Hamel}
\author[4]{Stavros Gaitanaros}
\author[1,5]{Nikolaos Bouklas\corref{cor1}}
\ead {nbouklas@cornell.edu}

\cortext[cor1]{Corresponding Author}

\address[1]{Sibley School of Mechanical and Aerospace Engineering, Cornell University, Ithaca, NY, USA\fnref{label1}}
\address[2]{Department of Aerospace \& Mechanical Engineering, University of Southern California, Los Angeles, 90007, CA, USA}
\address[3]{Sandia National Laboratories, Albuquerque, NM, USA\fnref{label2}}
\address[4]{DTU Engineering Technology, Technical University of Denmark, Ballerup, 2750 Denmark}
\address[5]{Pasteur Labs, Brooklyn, NY, USA\fnref{label5}}






\begin{abstract}
Architected metamaterials like foams and lattices exhibit complex responses governed by microstructural instabilities, localization, and phase-transition-like phenomena. Their behavior is further affected by heterogeneities inherent in their microstructure often caused through manufacturing processes. In this study we extend a gradient-enhanced, nonlocal continuum formulation to incorporate stochastic material heterogeneity through Gaussian random fields imposed on selected constitutive parameters. The framework enables independent control of both the amplitude and spatial correlation of material fluctuations while preserving thermodynamic consistency and regularization of localization. It also introduces a characteristic lengthscale ratio between the nonlocal and correlation lengthscales, that enables modeling at the limit of random or spatially correlated microstructures. Finite element simulations of confined compression and indentation show that heterogeneity fundamentally alters phase nucleation, localization morphology, and macroscopic response. Overall, the proposed framework provides a unified approach for linking stochastic material variability to instability-driven mechanics in architected metamaterials, enabling improved understanding of imperfection sensitivity, stability and design. It showcases how heterogeneity alone can influence characteristic features of the response, such as stability, slope of the plateau region, and elimination of the initial elastic regime. 
 \end{abstract}


\end{frontmatter}

\section{Introduction}

Architected metamaterials are a broad class of materials with low densities and complex meso-structures, that are naturally arising or by-design, leading to unconventional emergent properties and responses. Architected materials like foams, microlattices, and origami structures, offer unusual combinations of lightweightness, stiffness, strength, and programmable mechanical responses, making them attractive for energy absorption, soft robotics, and tunable structural components \cite{bertoldi_flexible_2017, lin_folding_2020, joe_development_2020, zhao_modular_2025}. Under compression, low-density metamaterials frequently exhibit a characteristic sequence of regimes, namely an initial elastic response terminating at a limit load, an extended stress plateau, and densification at large deformations \cite{gibson_mechanical_2000, gong_compressive_2005, han_compressive_1998, andrews_compressive_1999, mills_high_1999}. For metallic and polymeric foams, this macroscopic behavior is associated with underlying mesostructural mechanisms involving cell-wall or ligament bending/buckling, plastic yielding, formation of bands of collapsed cells, and progressive compaction \cite{deshpande_isotropic_2000, bastawros_experimental_2000, bardenhagen_insight_2005, schuler_deformation_2013}. The same features-stress plateaus, localization and sudden load drops-also arise in ordered truss- and plate-based metamaterials as well as in origami-based systems, where geometric nonlinearity and multistability can yield sharp transitions and sensitivity to imperfections \cite{liu_big_2020, rapaka_effect_2020, sang_tuneable_2024, glaesener2023predicting}. Depending on the properties of the base material as well as the meso-architecture, these materials can exhibit a wide range of mechanical behavior. Taking foams as an example, they can range from purely elastic and viscoelastic responses that fully recover \cite{gong2005compressive, hilyard2012low}, to brittle elastic behavior that experience extensive damage in compression \cite{chen2022strength}, as well as elastoplastic and viscoplastic responses exhibiting permanent deformations. 

The literature has focused on both order as well as disorder in this general materials class, leading to guiding principles for design, highlighting the corresponding benefits and design targets \cite{deshpande2001effective, wegst2015bioinspired}. With the design space at the microstructure level being inherently high-dimensional, the map from microstructure to macroscopic response is not simple, and as such, most of the aforementioned design principles come with constraining assumptions and restrictions regarding their validity. The relevance of order and disorder is also closely-tied to the corresponding manufacturing process. Additive manufacturing is often utilized for small scale samples, but more prominently foaming processes enable manufacturing of larger scale panels. These complex manufacturing processes introduce heterogeneities at different length-scales, e.g. variability of microscopic features such as strut-thickness, anisotropy, material concentrations at the junctions, and density-gradients at the specimen level \cite{bi2020additive, zheng2016multiscale}. There are occurrences where this variability leads to randomness (such as in the case of random foams \cite{roland2026physically}) but also others where the noise is spatially correlated (such as Torquato et al. \cite{torquato2002random}). In physical systems, it is very hard to control the degree and character of heterogeneity, and as such, theoretical and computational approaches aim to uncover the mechanistic principles that govern the complex response of these materials. But even for purely elastic architected metamaterials --which is the focus of the present work--, the influence of heterogeneity at the microstructural level is hard to quantify, especially in conjunction with its effect on the macroscopic response. This is because the microscopic and macroscopic elastic instabilities, that govern the metamaterial nonlinear mechanics, can be very sensitive to the variability of the micro-architecture.

Extensive theoretical and computational works have attempted to capture the macroscopic response of architected metamaterials focusing on the development of constitutive relations. These works include plasticity constitutive models for metallic foams \cite{deshpande_isotropic_2000,deshpande2001multi,chen2002size, yang2020continuum}, high-strain compressive descriptions for closed-cell foams \cite{mills_high_1999}, dynamic or crash and rate-dependent formulations \cite{zhang_constitutive_1998, sorrentino_simple_2007, koumlis_strain_2019, markert_biphasic_2008}, and soft elastic lattices \cite{khajehtourian2021continuum,khajehtourian2021soft} without considering the influence of microarchitecture heterogeneity. Discrete approaches, where the microstructure is explicitly discretized, \cite{gaitanaros2012crushing,gaitanaros2015effect,hooshmand2022mechanically,hooshmand2024m,GALLUP2026105137}, can provide critical information but are challenging to upscale to the structural level due to computational requirements. In multiscale approaches \cite{viot_multiscale_2008, miehe2007multiscale, geers2017homogenization, fritzen2015nonlinear} the underlying assumptions of periodicity that are commonly used, are often limiting the accurate prediction of localization. Gradient-enhanced, micromorphic and generalized continuum models allow exploration in this direction~\cite{neff2007geometrically, combescure2023selecting, iltchev2015computational} including a more recent array of works \cite{sperling2023enriched,guo2025reduced,maraghechi2024harvesting,van2020newton}. 

Structural defects, missing or partially coupled cells, and specimen-to-specimen variability can strongly influence collapse initiation and the development of deformation bands in cellular solids \cite{jeon_effect_2005, andrews_compressive_1999, jeon_cell_2009}. In architected systems, even small geometric imperfections can have outsized effects on stiffness, foldability, and nonlinear response \cite{liu_big_2020, rapaka_effect_2020}. Recent studies continue to emphasize that mechanical response is shaped by coupled sources of heterogeneity-material, geometric, and processing-induced-and that such heterogeneity influences localization patterns and stability under compression \cite{duan_quasi-static_2020, sang_tuneable_2024, yao_mechanical_2025}. In addition, classical energy-absorption characterizations highlight the importance of capturing plateau characteristics and densification onset accurately for design and selection \cite{avalle_characterization_2001}. 

Very often, constitutive models for architected metamaterials more broadly, are obtained directly from the macroscopic response \cite{landauer_experimental_2019,tao2021characterization, zheng2024hypercan, fernandez2021anisotropic, fernandez2022material}, inherently assuming that the deformation is homogeneous in the macroscale.
On the other hand, full-field imaging experiments demonstrate that deformation can be strongly heterogeneous: digital image correlation (DIC) and related full-field measurements reveal evolving strain textures, the emergence of localized bands, and distributed collapse events during compression \cite{wang_full-field_2002,  pierron_identification_2010,landauer_experimental_2019}. These observations support the interpretation of collapse as a phase transition from a rare to a dense phase, in which deformation localizes and then the phase-front propagates through the specimen \cite{wang_full-field_2002, gioia_energetics_2001}. 

In our recent work, we introduced a thermodynamically consistent, gradient-enhanced nonlocal continuum framework that captures instability-driven phase transitions in architected metamaterials-including localization and densification fronts, hysteresis in metastable and bistable regimes, and imperfection-insensitive macroscopic response-without explicitly resolving the underlying microstructure \cite{joshi2026instabilities}. In this framework a nonlocal lengthscale arises, which controls the thickness of the transition regions between rare and dense phases. That work did not focus on calibration towards a specific system, but rather on the requirements for the development of a model that captures common features that architected materials exhibit in their macroscopic response and in their microscopic deformation patterns. As such, it provides a general framework that can be specialized to explore specific features of interest. That work focused on samples of homogeneous material properties or graded material properties in simple loading scenarios; the latter was an elementary approach towards exploring the influence of heterogeneity in these materials systems.

The present work utilizes the aforementioned continuum-scale framework to study how stochastic heterogeneity alters localization patterns, macroscopic response, and stability. We represent heterogeneity through spatially correlated fluctuations using Gaussian Random Fields (GRF), imposed on constitutive parameters that govern elastic, volumetric, and transition-driving energetic contributions (the latter were previously shown to control phase transitions and corresponding instabilities), enabling controlled variation in both amplitude of the fluctuations  and in the correlation length. This allows us to directly interrogate how heterogeneity alters phase nucleation sites, promotes distributed versus band-like collapse patterns influencing rigidity percolation, and modifies the transition-like plateau trends. Such effects are consistent with experimental observations for evolution of heterogeneous deformation fields and collapse mechanisms in foams \cite{wang_full-field_2002, bastawros_experimental_2000, bardenhagen_insight_2005, sun_variation_2016}, and can also enable resolving questions that arise from controlled numerical exploration in perfect and imperfect microarchitectures \cite{hooshmand2022mechanically}. The numerical experiments focus on confined compression and indentation, to highlight the effect of concentrations in loading. Finally, because our simulations involve large deformation and contact during collapse, we also leverage robust finite element contact regularization concepts suitable for extreme deformation regimes using recent developments in third medium contact \cite{wriggers2025third}.  The framework allows to showcase how heterogeneity alone can influence characteristic features of the response, such as stability, slope of the plateau region, and elimination of the initial elastic regime. 


The remainder of the paper is organized as follows, Section \ref{Section:Nonlinear_Theory} summarizes the thermodynamically consistent gradient-enhanced continuum formulation that underpins the instability-driven volumetric transitions in isotropic architected metamaterials. Section \ref{Section:GRF_Implementation} introduces the stochastic representation of spatial heterogeneity via GRFs and describes how correlated fluctuations are mapped onto selected constitutive parameters. Section \ref{Section:FE_Implementation} presents the finite element implementation, including discretization, mixed-space construction, nonlinear solution strategy, and the treatment of large-deformation contact. Section \ref{Section:Results} reports numerical results: we first visualize representative heterogeneous parameter fields, then introduce a deterministic baseline gradation, and finally quantify how stochastic heterogeneity (amplitude and correlation length) alters localization morphology, front evolution, and macroscopic stability under confined compression and indentation. Section \ref{Section:Conclusion} concludes with a summary of the main findings and an outlook on extensions, with supplementary details in the Appendix.

\section{Nonlinear Theory}\label{Section:Nonlinear_Theory}

The nonlinear continuum framework adopted in this work builds directly on our previous gradient-enhanced formulation developed in Joshi et al. \cite{joshi2026instabilities}. Accordingly, only a concise summary of the governing kinematics, balance laws, and constitutive structure is provided here, with emphasis on aspects relevant to the present study. The theory is formulated for finite strains under isothermal and quasi-static conditions and is derived in a variational setting via the principle of virtual power; further, it is specialized for isotropy. In addition to the displacement field $\mathbf{u}$, a nonlocal volumetric internal variable $\tilde{J}$, and corresponding nonlocal lengthscale $\ell^{\text{nl}}$ is introduced to regularize volumetric localization. Dissipation enters the formulation exclusively through a viscous microforce associated with $\tilde{J}$, which is shown below to be fully consistent with the Clausius--Duhem inequality. This term makes the response rate-dependent but is only associated with volumetric deformations related to the phase transition from a rare to a dense phase. 

\subsection{Kinematics}

We consider a deformable continuum body $\mathcal{B}$ occupying a reference configuration $\Omega_0 \subset \mathbb{R}^3$ and a current configuration $\Omega \subset \mathbb{R}^3$ at time $t$. The motion $\mathbf{\varphi}: \Omega_0 \rightarrow \Omega$ maps a material point $\mathbf{X} \in \Omega_0$ to its spatial position $\mathbf{x} = \mathbf{\varphi}(\mathbf{X}, t) \in \Omega$. The displacement field is defined as 
\begin{equation}
    \mathbf{u}(\mathbf{X}, t) = \mathbf{\varphi}(\mathbf{X}, t) - \mathbf{X}.
\end{equation}
Local deformation is characterized by the deformation gradient
\begin{equation}
    \mathbf{F} = \nabla_{\mathbf{X}}\mathbf{\varphi}, \qquad J = \det \mathbf{F} > 0,
\end{equation}
where $J$ denotes the local volume ratio.

The right Cauchy--Green deformation tensor $\mathbf{C} = \mathbf{F}^T\mathbf{F}$ admits the principal invariants
\begin{equation}
    I_1 = \mathrm{tr}\,\mathbf{C}, \qquad I_2 = \frac{1}{2}\left[(\mathrm{tr}\,\mathbf{C})^2 - \mathrm{tr}(\mathbf{C}^2)\right], \qquad I_3 = \det \mathbf{C} = J^2.
\end{equation}
Throughout this work, the reference configuration is assumed stress-free, and the motion is smooth, invertible, and orientation-preserving. 

\subsection{Nonlocal volumetric description and governings}

Following Joshi et al. \cite{joshi2026instabilities}, we augment the classical kinematic description with a nonlocal volumetric internal variable $\tilde{J}$, which serves as a spatial averaging of the local volume ratio $J$. The resulting differential nonlocal formulation can be interpreted as the gradient-enhanced limit of integral-type averaging in the sense of Ba\v{z}ant \cite{bavzant2002nonlocal, pijaudier1987nonlocal}. 

The governing equations are derived in a variational setting using the principle of virtual power. The internal mechanical power in the reference configuration is written as
\begin{equation}
    \mathcal{P}_{\mathrm{int}} = \int_{\Omega_0}\left(\mathbf{P}:\nabla \dot{\mathbf{u}} + f_{\tilde{J}}\dot{\tilde{J}} + \boldsymbol{\xi}_{\tilde{J}}\cdot\nabla\dot{\tilde{J}}\right)\,\mathrm{d}V,
\end{equation}
where $\mathbf{P}$ is the first Piola--Kirchoff stress, $f_{\tilde{J}}$ is the scalar microforce conjugate to $\tilde{J}$, and $\boldsymbol{\xi}_{\tilde{J}}$ is the higher-order microstress conjugate to $\nabla\tilde{J}$. The generalized forces are additively decomposed into equilibrium and dissipative contributions,
\begin{equation}\label{Eq:Decomp_Forces}
    \mathbf{P} = \mathbf{P}^{\mathrm{eq}} + \mathbf{P}^{\mathrm{visc}}, \qquad f_{\tilde{J}} = f_{\tilde{J}}^{\mathrm{eq}} + f_{\tilde{J}}^{\mathrm{visc}}, \qquad \boldsymbol{\xi}_{\tilde{J}} = \boldsymbol{\xi}_{\tilde{J}}^{\mathrm{eq}} + \boldsymbol{\xi}_{\tilde{J}}^{\mathrm{visc}}.
\end{equation}
In the present formulation, we do not consider general viscoelasticity of the solid matrix. Instead, dissipation is introduced only through the evolution of the nonlocal volumetric measure $\tilde{J}$, which is used to regularize phase transition. Accordingly, the macroscopic stress and higher-order microstress are taken to be purely energetic, so that
\begin{equation}
    \mathbf{P}^{\mathrm{visc}} = 0, \qquad \boldsymbol{\xi}_{\tilde{J}}^{\mathrm{visc}} = 0,
\end{equation}
while $f_{\tilde{J}}^{\mathrm{visc}}$ remains as the only dissipative contribution. 



Neglecting body forces, the external mechanical power, $\mathcal{P}_{\mathrm{ext}}$ over the reference configuration $\Omega_0$ is given by,
\begin{equation}
    \mathcal{P}_{\mathrm{ext}} = \int_{\partial\Omega_0} \mathbf{T}\cdot\dot{\mathbf{u}} \,\mathrm{d}A,
\end{equation}
where $\mathbf{T}$ denotes the prescribed traction on the boundary and $\dot{\mathbf{u}}$ is the material velocity. The surface measure $\mathrm{d}A$ is taken with respect to the reference configuration. Application of the principle of virtual power, requiring equality of internal and external power for all admissible variations, i.e., $\delta \mathcal{P}_{\mathrm{int}} = \delta \mathcal{P}_{\mathrm{ext}}$, yields the governing field equations in the reference configuration.

\begin{subequations}\label{Eq:Governing_Eqn}
    \begin{align}
        \nabla_{\mathbf{X}}\cdot\mathbf{P} = \mathbf{0} && \text{in } \Omega_0, \\
        f_{\tilde{J}}^{\mathrm{eq}} + \nabla_{\mathbf{X}}\cdot\boldsymbol{\xi}_{\tilde{J}} + f_{\tilde{J}}^{\mathrm{visc}} = 0 && \text{in } \Omega_0.
    \end{align}
\end{subequations}

The boundary $\partial\Omega_0$ is partitioned into standard and microstructural portions. Essential and natural boundary conditions are prescribed as
\begin{subequations}\label{Eq:BCs}
    \begin{align}
        \mathbf{u} = \check{\mathbf{u}} && \text{on } \partial\Omega_0^{u}, \\
        \mathbf{P}\cdot\mathbf{N} = \check{\mathbf{T}} && \text{on } \partial\Omega_0^{t}, \\
        \tilde{J} = \check{\tilde{J}} && \text{on } \partial\Omega_0^{\tilde{J}}, \\
        \boldsymbol{\xi}_{\tilde{J}}\cdot\mathbf{N} = \check{\iota} && \text{on } \partial\Omega_0^{\xi},
    \end{align}
\end{subequations}
where $\mathbf{N}$ denotes the outward unit normal in the reference configuration. Unless otherwise stated, homogeneous micro-traction $\check{\iota} = 0$ is imposed on free surfaces. 

\subsection{Helmholtz free energy density}

The Helmholtz free energy per unit reference volume is taken as a function of the local deformation, the nonlocal volumetric variable, and its gradient:

\begin{equation}
    \Psi = \Psi(\mathbf{F},\tilde{J}, \nabla\tilde{J}).
\end{equation}
For isotropic architected metamaterials, we adopt
\begin{equation}\label{Eq:Free_Energy_Density}
    \Psi = \underbrace{\frac{\mu}{2}(I_1 - 3 - 2\ln J) + \frac{\kappa}{2}(\ln J)^2}_{\Psi_{\mathrm{C}}} + \underbrace{\frac{\alpha}{2}\left(\frac{(1 - \tilde{J})^2}{2} - \beta(1 - \tilde{J})\right)^2}_{\Psi_{\mathrm{NC}}} + \underbrace{c(J - \tilde{J})^2}_{\Psi^{\mathrm{coup}}} + \underbrace{d\ell^{\mathrm{nl}^2}\|\nabla\tilde{J}\|^2}_{\Psi^{\mathrm{grad}}},
\end{equation}
where the four groupings of terms correspond to the (poly)convex, non-(poly)convex, coupling and gradient contributions to the free energy density. Where $\mu$ and $\kappa$ are shear and bulk moduli for the compressible Neo-Hookean model that is chosen for the polyconvex local contributions. $\alpha$ and $\beta$ control the non-convex energy landscape directly controlling phase transitions, $c > 0$ enforces consistency between $J$ and $\tilde{J}$, and $\ell^{\mathrm{nl}}$ introduces an intrinsic interaction length scale. 

\subsection{Thermodynamic consistency and dissipation potential}

The rate of change of the Helmholtz free energy density is given by
\begin{equation}
    \dot{\Psi} = \frac{\partial\Psi}{\partial\mathbf{F}}:\dot{\mathbf{F}} + \frac{\partial\Psi}{\partial\tilde{J}}\dot{\tilde{J}} + \frac{\partial\Psi}{\partial\nabla\tilde{J}}\cdot\nabla\dot{\tilde{J}}
\end{equation}
Combining this expression with the internal power and invoking the Clausius--Duhem inequality yields the dissipation inequality
\begin{equation}
    \mathcal{D} = \left(\mathbf{P} - \frac{\partial\Psi}{\partial\mathbf{F}}\right):\dot{\mathbf{F}} + \left(f_{\tilde{J}}^{\mathrm{eq}} - \frac{\partial\Psi}{\partial\tilde{J}}\right)\dot{\tilde{J}} + \left(\boldsymbol{\xi}_{\tilde{J}} - \frac{\partial\Psi}{\partial\nabla\tilde{J}}\right)\cdot\nabla\dot{\tilde{J}} + f_{\tilde{J}}^{\mathrm{visc}}\dot{\tilde{J}} \ge 0.
\end{equation}
The inequality is satisfied for all admissible processes by adopting the constitutive relations
\begin{equation}
    \mathbf{P} = \frac{\partial\Psi}{\partial\mathbf{F}}, \qquad f_{\tilde{J}} = \frac{\partial\Psi}{\partial\tilde{J}}, \qquad \boldsymbol{\xi}_{\tilde{J}} = \frac{\partial\Psi}{\partial\nabla\tilde{J}}.
\end{equation}
The remaining dissipation reduces to
\begin{equation}
    \mathcal{D} = f_{\tilde{J}}^{\mathrm{visc}}\dot{\tilde{J}} \ge 0.
\end{equation}
To ensure non-negative dissipation for all admissible rates $\dot{\tilde{J}}$, we introduce, following Gurtin \cite{gurtin1996generalized}, a quadratic dissipation potential of the form
\begin{equation}
    \mathcal{R}(\dot{\tilde{J}}) = \frac{1}{2}\eta\dot{\tilde{J}}^2, \quad \forall \; \eta \ge 0,
\end{equation}
where $\eta$ is a viscosity parameter associated with the phase transition\footnote{Such viscous regularization is often referred to as artificial viscosity in numerical implementations which was proposed in~\cite{richtmyer1948proposed} and popularized in~\cite{vonneumann_richmyer}. This approach is typically used in hydrodynamics calculations for shock capturing but has also been used in phase-field formulations \cite{miehe2010phase}.}, which gives rise to viscous microforces of generalized standard material form:
\begin{equation}
    f_{\tilde{J}}^{\mathrm{visc}} = \frac{\partial \mathcal{R}}{\partial \dot{\tilde{J}}} = \eta\dot{\tilde{J}}.
\end{equation}
The local dissipation density thus reduces to
\begin{equation}
    \mathcal{D} = \eta\dot{\tilde{J}}^2 \ge 0,
\end{equation}
thereby ensuring thermodynamic admissibility. Substituting the constitutive relations into the microforce balance in Eq. \ref{Eq:Governing_Eqn}(b) yields
\begin{equation}
    \frac{\partial \Psi}{\partial\tilde{J}} + \nabla_{\mathbf{X}}\cdot\left(\frac{\partial \Psi}{\partial\nabla\tilde{J}}\right) + \eta\dot{\tilde{J}} = 0.
\end{equation}

\subsection{Work-conjugate quantities}

The equilibrium work-conjugate quantities entering the governing equations follow directly from Eq. \ref{Eq:Free_Energy_Density}:
\begin{subequations}
    \begin{align}
        \mathbf{P} = \mu(\mathbf{F} - \mathbf{F}^{-T}) + \kappa\ln J\mathbf{F}^{-T} + 2c(J - \tilde{J})J\mathbf{F}^{-T}, \\
        f_{\tilde{J}} = \alpha\left[\frac{(1 - \tilde{J})^2}{2} - \beta(1 - \tilde{J})\right](\tilde{J} - 1 + \beta) - 2c(J - \tilde{J}), \\
        \boldsymbol{\xi}_{\tilde{J}} = 2d\ell^{\mathrm{nl^2}}\nabla\tilde{J}.
    \end{align}
\end{subequations}
Together with Eq. \ref{Eq:Governing_Eqn} and Eq. \ref{Eq:BCs}, these relations close the coupled macro--micro boundary value problem in a thermodynamically consistent manner.

\section{Gaussian Random Fields}\label{Section:GRF_Implementation}

The nonlinear continuum framework developed in Sec. \ref{Section:Nonlinear_Theory} defines the governing kinematics, balance laws, and constitutive response of the architected metamaterial in a deterministic setting. In the present section, this framework is extended to incorporate spatial material heterogeneity by allowing selected constitutive parameters in the Helmholtz free energy density to vary smoothly over the reference configuration. Importantly, the introduction of heterogeneity does not modify the underlying variational structure, governing equations, or thermodynamic consistency of the model; instead, randomness enters exclusively through parametric modulation of the material coefficients, which occurs at the local level. 

Regular architected metamaterials, despite the fact that they are typically modeled as periodic, exhibit microstructural imperfections and geometric irregularities that result in local variations in stiffness, compressibility, and instability thresholds. Such heterogeneity may arise from manufacturing tolerances, localized defects, or intrinsic variability within the constituent phases.  Depending on the intensity and character of the local variability, some architected metamaterials are considered as random, whereas others have manufacturing-induced long-range spatial correlations. Capturing these spatial fluctuations is therefore essential for accurately predicting the onset and evolution of localization, phase transitions, and instability. To this end, we introduce spatial randomness in the constitutive parameters of Eq. \ref{Eq:Free_Energy_Density} using Gaussian Random Fields (GRFs), which provide a statistically consistent framework for generating smooth, spatially correlated variations of material properties.

\subsection{Description of spatial heterogeneity}

Let $G(\mathbf{X})$ denote a zero-mean, stationary Gaussian random field defined on the reference configuration $\Omega_0$. The field is characterized by a prescribed variance $\sigma^2$ and correlation length $\ell^{\mathrm{corr}}$, which respectively control the amplitude and spatial extent of material fluctuations. The second-order statistics of $G$ are fully described by its covariance function,

\begin{equation}
    \mathbb{E}\left[G(\mathbf{X}), G(\mathbf{Y})\right] = \mathcal{C}(\|\mathbf{X} - \mathbf{Y}\|).
\end{equation}
Here, $\mathbf{X}$ and $\mathbf{Y}$ denote two arbitrary material points in the reference configuration, and $\mathcal{C}(\cdot)$ is the two-point covariance function.

The Matérn kernel offers a flexible description of spatial correlations and allows direct control over the smoothness of the random field. Its smoothness parameter is taken as $\nu = n_d / 2$, where $n_d$ is the spatial dimension, yielding fields that are once mean-square differentiable and compatible with the gradient-enhanced continuum framework adopted in this work.

Following the Whittle--Matérn construction \cite{lindgren2011explicit, roininen2014whittle}, the GRF $G$ may equivalently be represented as the weak solution of the stochastic elliptic partial differential equation
\begin{equation}\label{Eq:GRF_SPDE}
    \left(\gamma \nabla^2 + \delta\mathbb{I}\right)G = \mathcal{W} \quad \text{in } \Omega_0,
\end{equation}
where $\mathcal{W}$ denotes spatial Gaussian white noise with zero mean and unit variance, $\nabla^2 = \nabla\cdot\nabla$ is the Laplacian operator, and $\mathbb{I}$ is the identity operator acting on scalar fields. The operator on the left-hand side acts as a linear spatial filter that transforms uncorrelated noise into a smooth random field with the desired correlation structure. 
The coefficients $\gamma$ and $\delta$ are related to the target variance $\sigma^2$ and correlation length $\ell^{\mathrm{corr}}$ through
\begin{equation}
    \Lambda = \frac{\sqrt{8\nu}}{\ell^{\mathrm{corr}}}, \qquad \gamma = \frac{1}{\sigma\Lambda^{\nu}}\sqrt{\frac{\Gamma(\nu)}{(4\pi)^{n_d/2}}}, \qquad \delta = \gamma\Lambda^2,
\end{equation}
where $\Gamma(\nu)$ denotes the Gamma function. This parameterization ensures that the solution of Eq. \ref{Eq:GRF_SPDE} reproduces the prescribed Matérn covariance function and variance independently of spatial dimension. 

In this representation, the parameter $\Lambda$ sets the inverse correlation length of the field and therefore controls the spatial scale over which fluctuations are correlated: larger values of $\Lambda$ corresponds to shorter correlation lengths and more rapidly varying fields, while smaller values yield smoother, long-range correlated heterogeneity. The coefficient $\gamma$ scales the strength of the Laplacian smoothing operator and thus governs the overall roughness of the field, whereas $\delta$ acts as a zeroth-order regularization that balances diffusion and preserves the prescribed variance. Together, the operators $\gamma\nabla^2$ and $\delta\mathbb{I}$ define an elliptic filter that maps white noise into a Gaussian random field with controlled amplitude, smoothness, and spatial correlation.

\subsection{Stochastic sampling of material fields}
Independent realizations of the GRF are obtained by solving Eq. \ref{Eq:GRF_SPDE} for different realizations of the white noise forcing $\mathcal{W}$. Each realization yields a spatially correlated field $G_i(\mathbf{X})$, $i = 1,\dots,N_s$, defined over $\Omega_0$. The resulting ensemble of samples
\begin{equation}
    \left\{G_1(\mathbf{X}), \,G_2(\mathbf{X}), \dots, G_{N_s}(\mathbf{X})\right\}
\end{equation}
constitutes a statistically consistent representation of material heterogeneity with controlled variance and correlation length. The use of SPDE-based formulation ensures that the generated fields are mesh-consistent and remain well-defined under refinement, making the approach suitable for large-deformation finite element simulations \cite{tan2024scalable}. 

\subsection{Connection with the constitutive model}

Spatial heterogeneity is incorporated into the constitutive framework by modulating the material parameters appearing in the Helmholtz free energy density Eq. \ref{Eq:Free_Energy_Density}. For a generic constitutive parameter $\bar{p} \in \{\bar{\mu}, \bar{\kappa}, \bar{\alpha}, \bar{\beta}\}$, the corresponding heterogeneous field is defined as

\begin{equation}
    p(\mathbf{X}) = \bar{p}\exp\left(G_p(\mathbf{X}) - \frac{1}{2}\sigma_p^2\right),
\end{equation}
where $G_p(\mathbf{X})$ is a Gaussian random field realization with variance $\sigma_p^2 = \ln(1 + \mathrm{CV}^2)$, ensuring that $p(\mathbf{X})$ remains strictly positive and has a mean value $\bar{p}$. The coefficient of variation $\mathrm{CV}$ thus provides a direct and physically interpretable measure of heterogeneity intensity. 

Each constitutive parameter may be sampled independently or in combination, enabling controlled perturbations of the baseline material response without altering the governing balance laws or variational structure. In this way, spatial heterogeneity enters the model purely parametrically, preserving the thermodynamic consistency of the gradient-enhanced framework while providing a systematic means to probe imperfection sensitivity.

It also has to be noted that the correlation length $\ell^{\text{corr}}$ can be compared to the intrinsic nonlocal lengthscale $\ell^{\text{nl}}$. As the latter is part of a low-pass filter through the Helmholtz operator that appears in \ref{Eq:Governing_Eqn}(b) this introduces three regimes: i) $\ell^{\text{corr}}<\ell^{\text{nl}}$ where the correlated noise passes through a filter and is approximates white noise, as would be expected in a random medium, ii) $\ell^{\text{corr}}\approx \ell^{\text{nl}}$ which is a transition range, and iii) $\ell^{\text{corr}}>\ell^{\text{nl}}$ which allows resolved correlated structures at the continuum scale. Thus, controlling the ratio $\ell^{\text{corr}}/\ell^{\text{nl}}$ enables probing the response of architected metamaterials that have correlated structures versus ones which are approximately random.
\section{Finite Element Implementation} \label{Section:FE_Implementation}

The coupled macro--micro boundary value problem derived in Sec. \ref{Section:Nonlinear_Theory} is solved numerically within a mixed finite element setting. The formulation and solution strategy follow closely to the implementation presented in Joshi et al. \cite{joshi2026instabilities}. Importantly, the introduction of heterogeneity does not alter the governing equations or variational structure, and enters the numerical scheme exclusively through parametric evaluation of the constitutive response. 

The finite element discretization is implemented using the \texttt{FEniCS} computational framework \cite{logg2012automated, FEniCSAlnaes2015}, with weak forms expressed symbolically using the Unified Form Language (UFL) \cite{UFL}. This approach enables direct transcription of the variational statements derived in Sec. \ref{Section:Nonlinear_Theory} into a computationally consistent and differentiable finite element formulation. 

\paragraph{Function spaces and admissible variations.}
Let $\Omega_0 \subset \mathbb{R}^d$ denote the reference configuration, with $d$ being the spatial dimension. We define the trial spaces for the displacement field and the nonlocal volumetric variable as
\begin{subequations}\label{Eq:Function_Spaces and admissible variations}
    \begin{align}
        \mathbb{U} = \left\{\mathbf{u} \in [H^1(\Omega_0)]^d\;\big|\;\mathbf{u} = \check{\mathbf{u}} \;\text{on } \partial\Omega_0^u\right\}, \\
        \mathbb{J} = \left\{\tilde{J} \in H^1(\Omega_0)\;\big|\;\tilde{J} = \check{\tilde{J}} \;\text{on } \partial\Omega_0^{\tilde{J}}\right\}
    \end{align}
\end{subequations}
with associated admissible variations $(\delta\mathbf{u}, \delta\tilde{J}) \in \mathbb{U}\times\mathbb{J}$ vanishing on the corresponding Dirichlet boundaries. We note that $\tilde{J}$ is subjected to natural boundary condition corresponding to homogeneous micro-traction $\check{\iota} = 0$, on free surface.


\paragraph{Weak formulation.} The finite element problem is to find $(\mathbf{u}, \tilde{J}) \in \mathbb{U} \times \mathbb{J}$ such that, for all admissible variations $(\delta\mathbf{u}, \delta\tilde{J})$, 
\begin{subequations}\label{Eq:Weak_Form}
    \begin{align}
        \int_{\Omega_0}\mathbf{P}(\mathbf{F}, \tilde{J}) : \nabla\delta\mathbf{u}\,\mathrm{d}V = \int_{\partial\Omega_0^{t}}\check{\mathbf{T}}\cdot\delta\mathbf{u}\,\mathrm{d}A, \\
        \int_{\Omega_0}\left(f_{\tilde{J}}\,\delta\tilde{J} + \boldsymbol{\xi}_{\tilde{J}}\cdot\nabla\delta\tilde{J} + \eta\,\dot{\tilde{J}}\,\delta\tilde{J}\right)\mathrm{d}V = \int_{\Omega_0^{\xi}}\check{\iota}\,\delta\tilde{J}\,\mathrm{d}A
    \end{align}
\end{subequations}
Here, the stress $\mathbf{P}$, scalar microforce $f_{\tilde{J}}$, and higher-order micro-stress $\boldsymbol{\xi}_{\tilde{J}}$ are obtained directly from the Helmholtz free energy density Eq. \ref{Eq:Free_Energy_Density}, ensuring exact consistency with the thermodynamic framework. 

\subsection{Temporal discretization}

The viscous microforce term associated with $\tilde{J}$ is discretized in time using a backward Euler scheme. Denoting the time increment by $\Delta t$, the rate $\dot{\tilde{J}}$ is approximated as
\begin{equation}
    \dot{\tilde{J}} \approx \frac{\tilde{J}^{n+1} - \tilde{J}^n}{\Delta t}.
\end{equation}
It is noted that $\eta$ is chosen small enough to not significantly affect the stable part of the quasi-static response, and only influence the dissipation associated to phase transitions. 

\subsection{Spatial discretization}

The mixed finite element discretization employs continuous Lagrange elements, using a Taylor--Hood interpolation with quadratic shape functions for the displacement field and linear shape functions for the nonlocal volumetric variable,

\begin{equation}
    \mathbf{u} \in [\mathbb{P}_2]^d, \qquad \tilde{J} \in \mathbb{P}_1.
\end{equation}
This choice satisfies the regularity requirements imposed by the gradient term in the free energy and provides a stable and robust approximation of the coupled macro-micro problem. As demonstrated in Joshi et al. \cite{joshi2026instabilities}, this interpolation yields mesh-independent resolution of localization and phase transition fronts in gradient-enhanced continua. 

\subsection{Incorporation of spatial heterogeneity}

Spatial material heterogeneity is incorporated through parametric modulation of the constitutive coefficients appearing in the Helmholtz free energy density, Eq. \ref{Eq:Free_Energy_Density}, as described in Sec. \ref{Section:GRF_Implementation}. For each realization, the spatially correlated random fields $\mu(\mathbf{X})$, $\kappa(\mathbf{X})$, $\alpha(\mathbf{X})$, and $\beta(\mathbf{X})$ are evaluated pointwise at the quadrature points during assembly of the weak form. These heterogeneous fields enter the formulation exclusively through the local constitutive response, while the global finite element spaces, variational structure, and solution procedure remain unchanged. 
This construction ensures that material disorder is introduced in a consistent manner. The details of the numerical implementation used to generate these GRFs are provided in the Appendix \ref{appendix:Numerical_Impl_GRF}. 

\subsection{Nonlinear solution strategy}

The spatial and temporal discretization described above leads to a nonlinear system of algebraic equations for the coupled unknowns $(\mathbf{u}, \tilde{J})$ at each pseudo-time increment. The primary solution strategy is a monolithic Newton--Raphson method, in which the residual and consistent tangent operator are obtained by automatic differentiation of the weak form using UFL. The resulting linearized systems are solved using \texttt{PETSc} Krylov subspace methods with algebraic multi-grid (AMG) preconditioning. 
Convergence of the monolithic scheme is assessed based on the $L^2$-norm of the global residual, with a relative and absolute tolerance of $10^{-9}$ imposed on both the displacement and nonlocal volumetric fields. This stringent criterion ensures accurate resolution of the coupled macro-micro response, particularly in the presence of strong nonlinearity and non-convexity associated with volumetric phase transitions. 


In the majority of simulations, the coupled system is solved using a fully monolithic Newton--Raphson scheme, which provides quadratic convergence when the solution remains within the basin of attraction of nonlinear equilibrium. However, in parameter regimes characterized by strong non-convexity or the onset of sharp volumetric localization, the monolithic solver may fail to converge.
In such cases, the solution procedure switches to a staggered (alternate minimization) scheme, using the last available iterate of the monolithic solve as the initial condition. Within this fallback strategy, the displacement field $\mathbf{u}$ and the nonlocal volumetric variable $\tilde{J}$ are updated sequentially at fixed pseudo-time until convergence is recovered. The staggered iterations are terminated when the incremental change in the nonlocal field satisfies
\[
\|\tilde{J}^{k+1} - \tilde{J}^{k}\|_{L^\infty(\Omega_0)} < 10^{-3},
\]
which has been found sufficient to restore stability while preserving the correct qualitative and quantitative features of phase-front evolution. 

This adaptive monolithic-to-staggered solution strategy follows the approach established in Joshi et al. \cite{joshi2026instabilities} and provides a robust means of traversing non-convex energy landscapes, enabling reliable simulation of large-deformation instabilities and metastable phase transition in heterogeneous architected metamaterials. 


\section{Results and Discussion} \label{Section:Results}

The numerical experiments reported in this section are conducted using the stochastic, gradient-enhanced, mixed finite element framework introduced in Sec. \ref{Section:FE_Implementation}. Our objective is to quantify how spatial heterogeneity in the constitutive parameters reshapes (i) the macroscopic force--displacement response, (ii) the onset and morphology of densification and localization, and (iii) the stability of the transition regime under confined compression and indentation. The stochastic setting enables systematic control of both the heterogeneity amplitude, through the coefficient of variation (CV), and its characteristic length scale, through the correlation length ($\ell^{\mathrm{corr}}$), thereby isolating the role of correlated disorder from other sources of regularization.

All simulations consider a two-dimensional problems in plane strain conditions and quasi-static loading. Two boundary value problems are considered. First, confined compression is enforced via rollers on the lateral boundaries and the bottom edge, together with a frictionless rigid plate applying the displacement-controlled compression from the top. Second, indentation is performed using a rigid circular indenter in frictionless contact, enforced through the third-medium contact formulation summarized in Appendix \ref{appendix: AM_TMC_Indent}. In all cases, the mechanical response is characterized by the macroscopic force--displacement relation, normalized by the shear modulus $\mu$, together with representative snapshots of Jacobian field $J$, which serves as a direct indicator of volumetric densification and phase transition.

Heterogeneity is introduced through Gaussian Random Fields (GRFs) as described in Sec. \ref{Section:GRF_Implementation}. For each constitutive parameter $(p \in \{\mu, \kappa, \alpha, \beta\})$, a zero-mean, unit-variance GRF $(G_p(\mathbf{X}))$ is generated and mapped to a strictly positive, mean-preserving field using a normalized lognormal transform,
\begin{equation}\label{Eq:Lognormal_Map}
    G_p^+(\mathbf{X}) = \frac{\exp\!\big(\sigma_p\,G_p(\mathbf{X})\big)}{\left\langle\exp\!\big(\sigma_p\,G_p\big)\right\rangle_{\Omega_0}},
\end{equation}
where $\langle\cdot\rangle_{\Omega_0}$ denotes the spatial average over the reference domain $\Omega_0$. This construction guarantees $G_p^{+}(\mathbf{X}) > 0$ everywhere while preserving the prescribed spatial correlation structure. The parameter $\sigma_p$ is selected such as that the resulting positive field attains the target coefficient of variation. The heterogeneous constitutive parameter is then defined multiplicatively as
\begin{equation}\label{Eq:Param_Modulation}
    p(\mathbf{X}) = \bar{p}G_p^+(\mathbf{X})
\end{equation}
with $\bar{p}$ denoting the nominal (homogeneous) value.
Unless stated otherwise, the nominal parameter values are fixed as
\[
\mu = 1.0, \qquad \kappa = 1.0, \qquad \alpha = 150.0, \qquad \beta = 0.5,
\]
and these values serve as the mean parameters when heterogeneity is introduced through the GRFs. Normalization of the force--displacement response by $\mu$ allows direct comparison across all cases and isolates the influence of heterogeneity in the remaining parameters. 

Owing to the lognormal mapping, the heterogeneous fields are inherently skewed, with comparatively greater spatial support in locally compliant regions than in locally stiff extremes. This feature is advantageous for the present study: it avoids non-physical negative parameters while introducing smooth, spatially correlated perturbations of controlled intensity. Unless otherwise stated, stochastic fields are generated independently for each heterogeneous parameter while holding the remaining parameters uniform, allowing the influence of shear ($\mu$), volumetric ($\kappa$), and phase-transition related parameters ($\alpha, \beta$) disorder to be examined in isolation. We additionally consider cases in which all parameters are heterogeneous simultaneously to assess the combined effect of multi-parameter disorder on localization and macroscopic stability. For each stochastic configuration, multiple independent realizations are performed to distinguish reproducible trends from realization-specific variability. 


The remainder of this section is organized as follows. We first visualize representative heterogeneous fields through property realizations and then a simplified baseline gradation in $\kappa(\mathbf{X})$, which serves as a controlled reference for how property variations bias densification localization. Next, we present confined compression results with GRF-induced heterogeneity, including systematic variation of CV and $\ell^{\mathrm{corr}}$, to demonstrate how correlated disorder alters localization morphology and has the potential to stabilize the transition plateau. Finally, we examine indentation of heterogeneous specimens using third-medium contact, highlighting how indenter geometry and stochastic heterogeneity jointly govern the indentation response and densification patterns.

\subsection{GRF Realizations}
To illustrate the nature of the spatial heterogeneity introduced through the GRF formulation, we present representative realizations of the non-convex parameter $\alpha(\mathbf{X})$ over the computational domain. These plots serve to visualize how the prescribed statistical descriptors -- namely the coefficient of variation (CV) and the correlation length $\ell^{\text{corr}}$ -- manifest as smooth, spatially correlated fluctuations at the continuum scale. 
These realizations correspond to numerical solutions of stochastic elliptic equation, Eq. \ref{Eq:GRF_SPDE} introduced in Sec. \ref{Section:GRF_Implementation}, computed using the finite element framework in \texttt{FEniCS}, with implementation details provided in Appendix \ref{appendix:Numerical_Impl_GRF}.

\begin{figure}[h!]
    \centering
    \includegraphics[width=\linewidth]{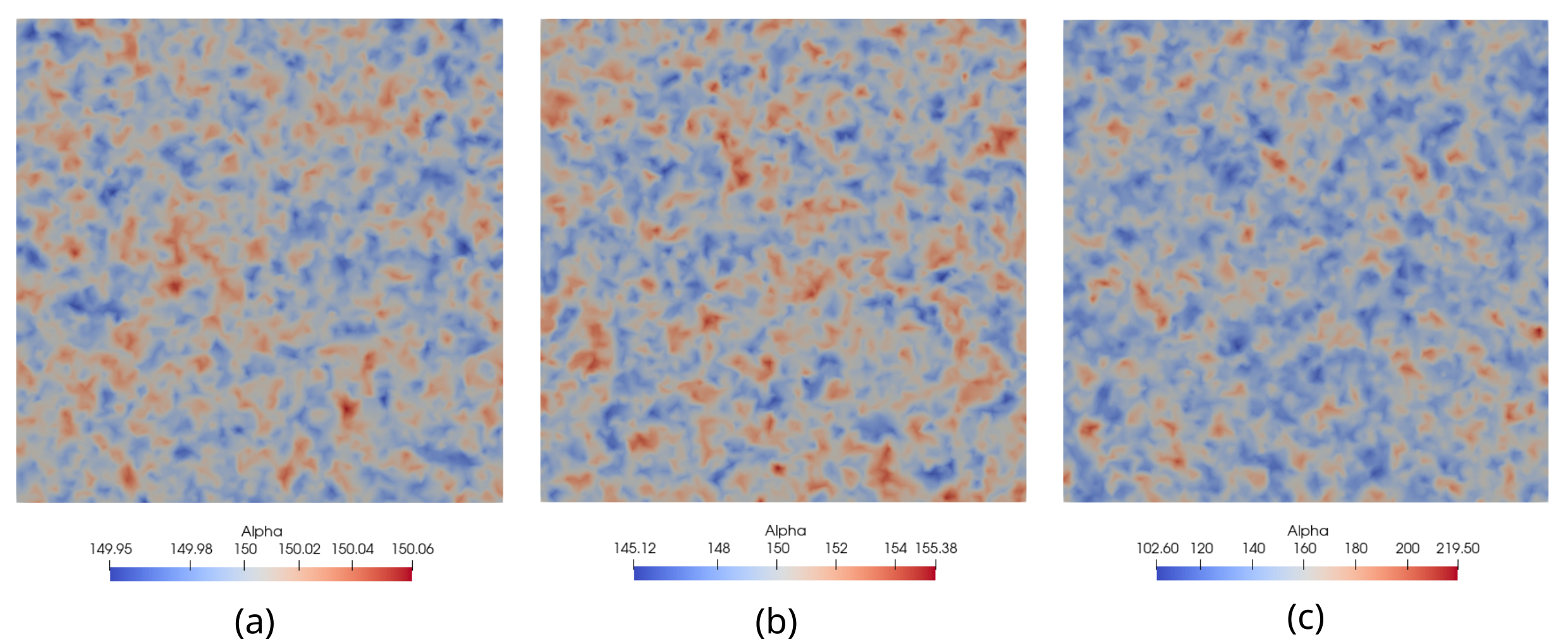}
    \caption{The realizations showcase the variation of $\alpha(\mathbf{X})$ with CV (a) $0.01\%$, (b) $1.0\%$, and (c) $10.0\%$ for an unstructured mesh with $\ell^{\text{corr}} = 4\ell^{\text{nl}}$.}
    \label{fig:alpha_CV_variation}
\end{figure}

Fig. \ref{fig:alpha_CV_variation} shows three realizations of $\alpha(\mathbf{X})$ generated on an unstructured mesh with fixed correlation length $\ell^{\text{corr}} = 4\ell^{\text{nl}}$\footnote{It is reminded, that the nonlocal lengthscale needs to be resolved by the mesh utilized in the numerical calculations.}, while systematically increasing the coefficient of variation from $0.01\%$ to $10.0\%$. At the lowest CV, the field is nearly uniform, with fluctuations that are visually negligible relative to the mean value, corresponding to an effectively homogeneous material. 
As the CV increases to $1.0\%$, localized regions of elevated and reduced $\alpha$ begin to emerge, while the field remains smooth and free of mesh-scale oscillations. For $\text{CV} = 10.0\%$, the heterogeneity becomes pronounced, with clearly identifiable soft and stiff regions distributed throughout the domain, reflecting substantial spatial variability that is expected to influence the non-convex energy landscape. 

\begin{figure}[h!]
    \centering
    \includegraphics[width=\linewidth]{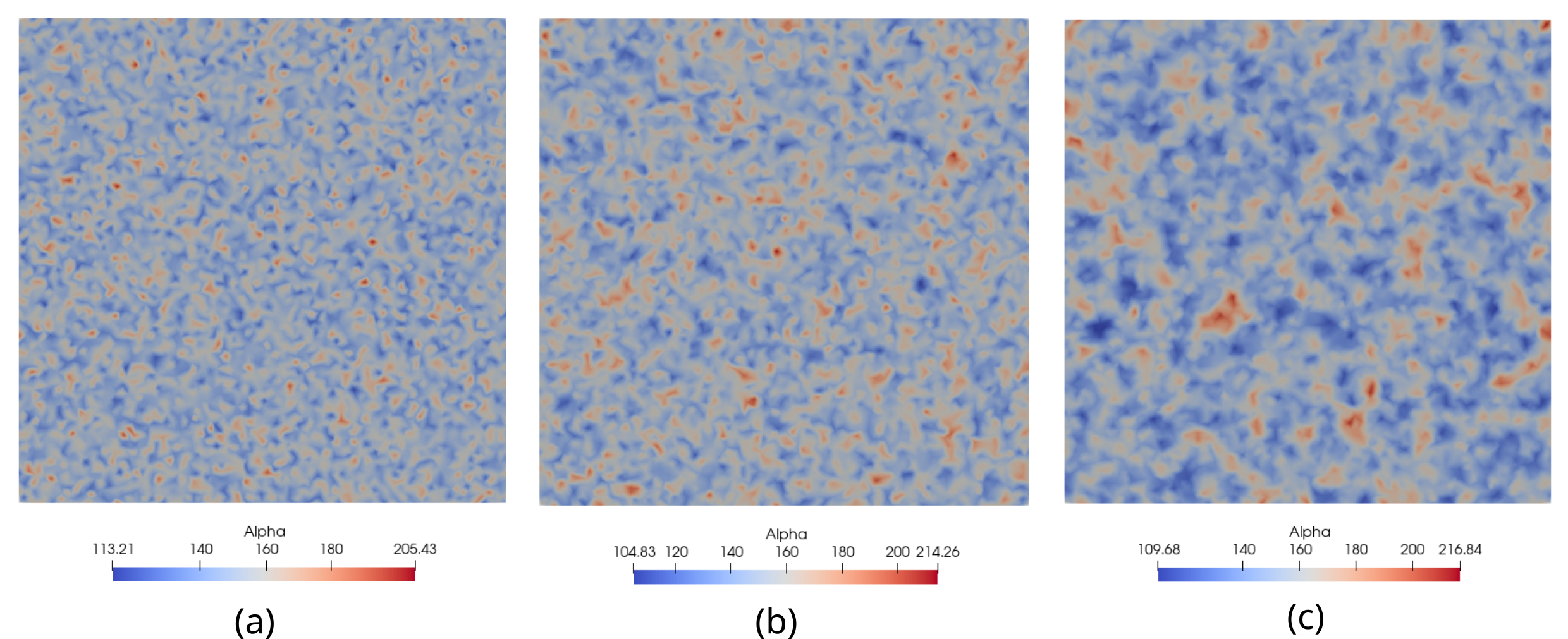}
    \caption{The realizations showcase the variation of $\alpha(\mathbf{X})$ with $\ell^{\text{corr}} =$ (a) $\ell^{\text{nl}}$, (b) $2\ell^{\text{nl}}$, and (c) $4\ell^{\text{nl}}$ for unstructured mesh with $\text{CV} = 10.0\%$.}
    \label{fig:alpha_Corr_variation}
\end{figure}

The influence of the correlation length is illustrated in Fig. \ref{fig:alpha_Corr_variation}, where the coefficient of variation is held fixed at $\text{CV} = 10.0\%$ and $\ell^{\text{corr}}$ is varied relative to the intrinsic non-local length scale $\ell^{\text{nl}}$. For $\ell^{\text{corr}} = \ell^{\text{nl}}$, the field exhibits rapid spatial fluctuations, with heterogeneity distributed over relatively short length scales. As the correlation length increases to $2\ell^{\text{nl}}$ and $4\ell^{\text{nl}}$, the spatial variations become progressively smoother, giving rise to larger coherent regions in which $\alpha(\mathbf{X})$ remains approximately uniform. This behavior is consistent with the underlying Matérn covariance structure and highlights the role of $\ell^{\text{corr}}$ in controlling the characteristic size of heterogeneous patches. 



\subsection{Baseline Gradation of $\kappa(\mathbf{X})$}
Before introducing stochastic heterogeneity through Gaussian Random Fields (GRFs), we consider a deterministic vertical gradation of the bulk modulus $\kappa(\mathbf{X})$ (of the Neo-Hookean model, utilized for the polyconvex part of the model) as a controlled baseline. 
The results shown here are adapted from Joshi et. al. \cite{joshi2026instabilities} in confined compression. They provide a baseline of more idealized cases (homogeneous material properties, and simple directional gradation), for comparison with the stochastic cases examined later. This is important as it allows to isolate effects that influence the model response that solely arises due to the constitutive model choice, structural (deterministic) heterogeneity, or the introduced stochasticity. This is achieved by comparing a background homogeneous material, simple directional gradation, and stochastic realizations of heterogeneity.
In this setting, $\kappa(\mathbf{X})$ varies monotonically along the loading direction, while all other constitutive parameters remain uniform. The bottom of the specimen is fixed at $\kappa = 1.0$, and a linear gradation is imposed such that the bulk modulus decreases toward the top. For the graded case, the top surface is prescribed to be $10.0\%$ softer than the bottom, whereas the homogeneous case corresponds to no spatial variation. Unlike the GRF-based heterogeneity introduced in the subsequent subsections, this graded field is deterministic and imposes a prescribed directional bias in volumetric stiffness.

\begin{figure}[h!]
    \centering
    \includegraphics[width=\linewidth]{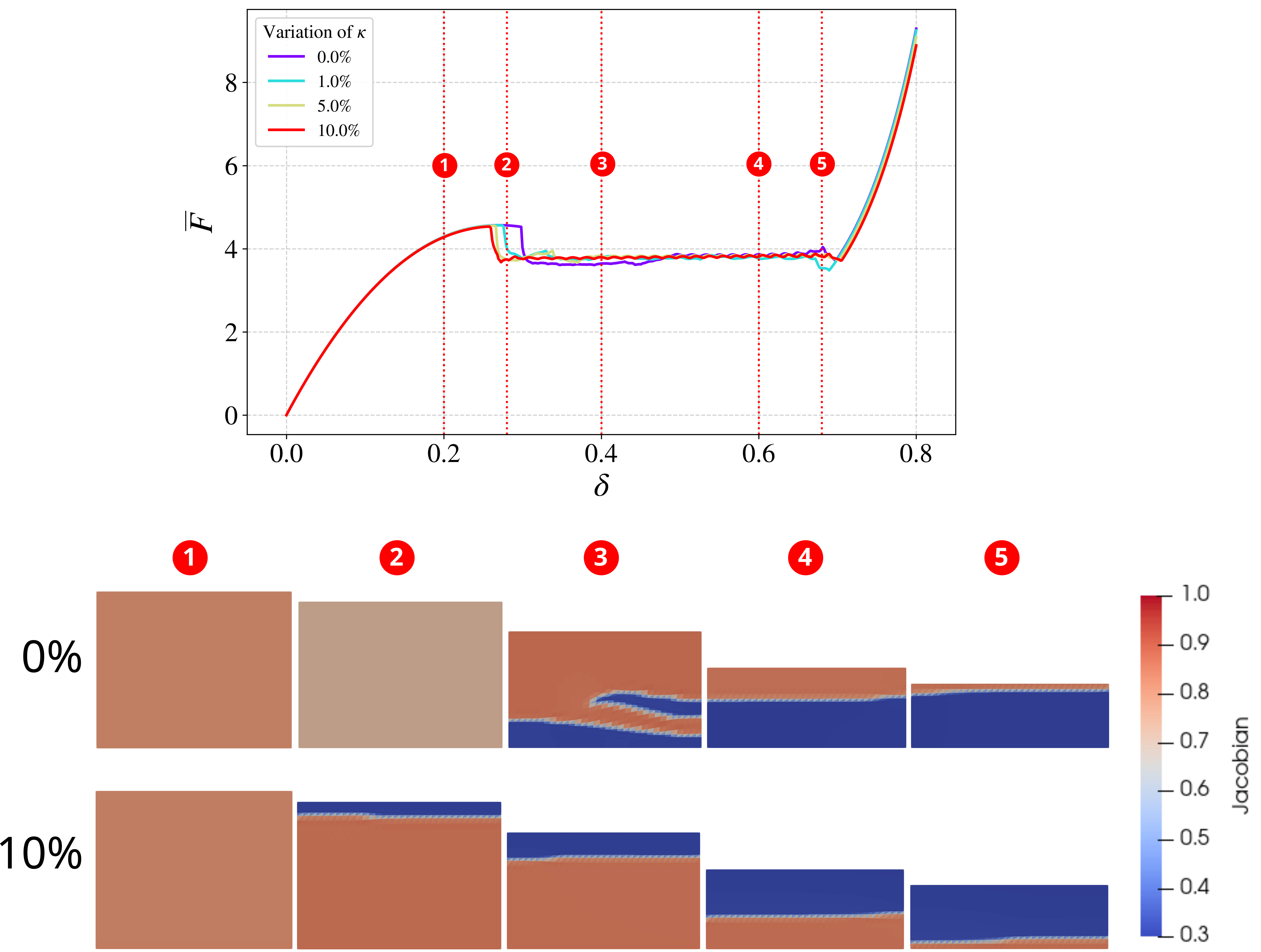}
    \caption{Adapted from Joshi et al.~\cite{joshi2026instabilities}. Influence of vertical grading in the bulk modulus $\kappa$ on the force–-displacement response and Jacobian field $J$ of a metastable material. Graded profiles correspond to $0.0\%$ and $10.0\%$ variation across the domain height. Snapshots for the $0.0\%$ and $10.0\%$ cases at five loading stages illustrate the impact of property heterogeneity on densification evolution.}
    \label{fig:kappa_var_gradient}
\end{figure}
Fig. \ref{fig:kappa_var_gradient} presents the corresponding force--displacement responses together with representative contours of the Jacobian field $J$ at five characteristic loading stages. In the homogeneous case ($0.0\%$ variation), the volumetric stiffness is spatially uniform and no material-preferred nucleation site exists. Consequently, the onset of localized compaction is governed by small numerical perturbations inherent to the discretization and solver. Localization therefore initiates at arbitrary locations, within the specimen and rapidly develops into a specimen-spanning band, followed by relatively uniform propagation with continued loading. In this context, the specimen-spanning band induces a simply connected region where loss of ``volumetric rigidity percolation''\footnote{The terminology ``volumetric rigidity percolation'', as used here, corresponds to the formation of an end-to-end band where a volumetric phase transition and corresponding instability can take place. It should not convey that volumetric resistance is permanently lost.} is instantaneously observed.
In contrast, for the strongly graded case ($10.0\%$ variation), the imposed stiffness contrast governs the response. The softer region near the top of the specimen becomes energetically favorable to undergo a phase transition, and densification nucleates exclusively within this compliant zone. The localization process is no longer influenced by numerical perturbations, and no additional nucleation sites are observed. An additional feature on both of these cases, is that prior to the plateau of the response, an instability is observed in the form of a limit load and subsequent load reduction. This feature is consistent over both of these cases that do not showcase stochastic variability.

These results demonstrate that for a perfectly homogeneous material multiple admissible localization paths exist following the instability and phase transition, but deterministic spatial variations in $\kappa(\mathbf{X})$ provides a robust mechanism for steering the nucleation and evolution of densified phase. 

This baseline study establishes a clear physical link between heterogeneity, localization and stability, which will be extended in the following subsections to stochastic heterogeneity introduced through GRFs.

\subsection{Heterogeneity in confined compression}\label{Subsection:CV_Var}
We next investigate the effect of stochastic material heterogeneity introduced through Gaussian Random Fields (GRFs) on the confined compression response. In contrast to the deterministic grading discussed earlier, heterogeneity is introduced selectively in the constitutive parameters of the free energy density, namely $\alpha(\mathbf{X})$, $\beta(\mathbf{X})$, $\kappa(\mathbf{X})$, and $\mu(\mathbf{X})$. This formulation enables a systematic assessment of how random spatial fluctuations in distinct energetic contributions influence phase nucleation, localization mechanisms, and macroscopic stability. 
\begin{figure}[h!]
    \centering
    \includegraphics[width=\linewidth]{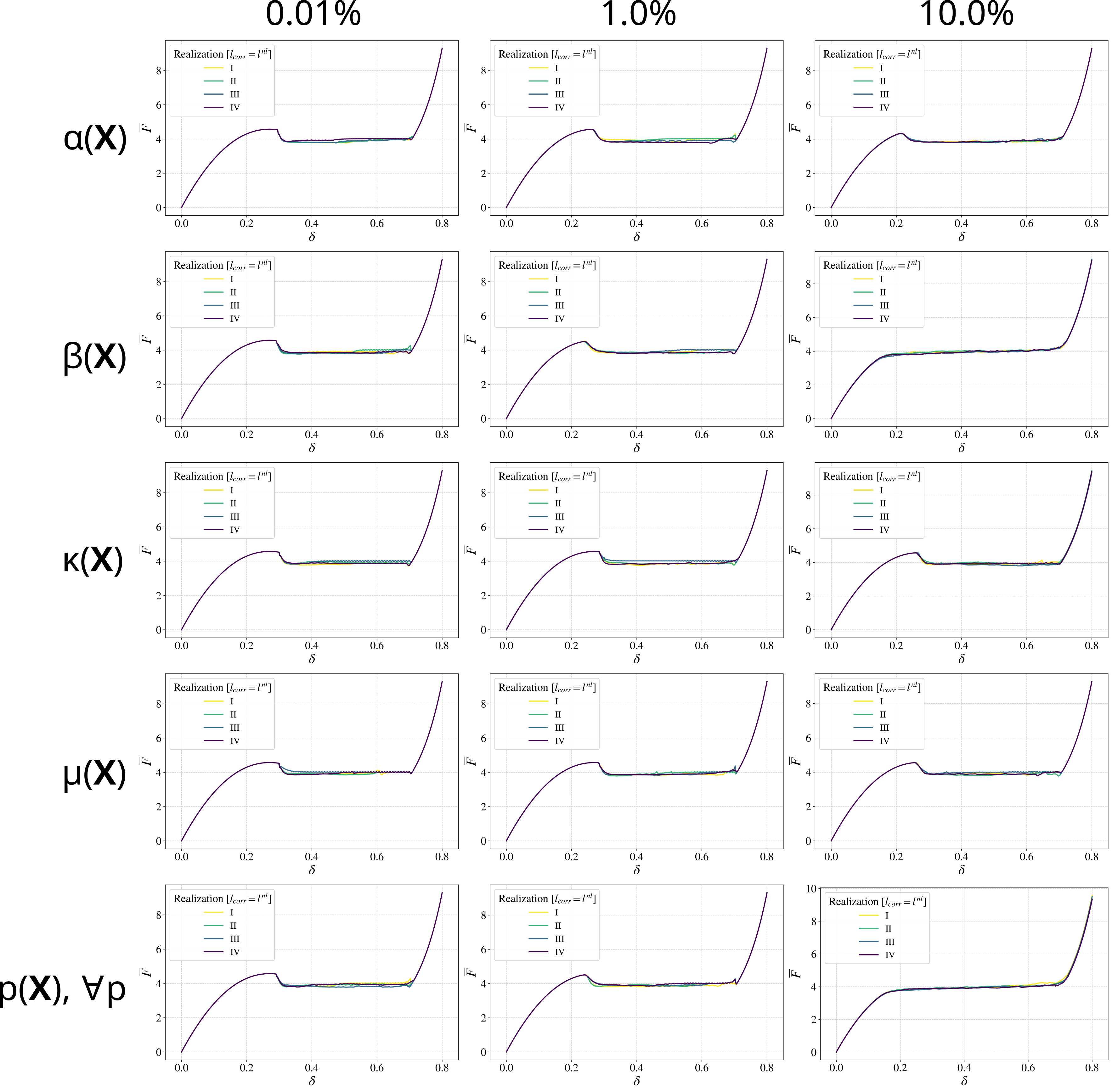}
    \caption{Influence of coefficient of variation of confined compression response for $\ell^{\mathrm{corr}} = \ell^{\mathrm{nl}}$. Force--displacement curves are shown for $\mathrm{CV} = 0.01\%$, $1.0\%$, and $10.0\%$, with four realizations for each case. Rows correspond to heterogeneity introduced individually in $\alpha$, $\beta$, $\kappa$ and $\mu$, followed by the case where all parameters are heterogeneous, $p \in \{\alpha, \beta, \kappa, \mu\}$, simultaneously. Increasing variance progressively stabilizes the transition plateau and mitigates limit-load drops, with the most pronounced stabilization observed when all parameters are heterogeneous at $\mathrm{CV} = 10.0\%$.}
    \label{fig:CV_triplets}
\end{figure}
Fig. \ref{fig:CV_triplets} examines the role of heterogeneity magnitude through a parametric study in which the coefficient of variation (CV) is increased at fixed correlation length $\ell^{\mathrm{corr}} = \ell^{\mathrm{nl}}$. For each case, four independent realizations are shown, demonstrating the robustness of the observed trends. When heterogeneity is introduced only in $\alpha$, increasing $\mathrm{CV}$ progressively reduces the severity of the limit-load drop, indicating that spatial variability in the depth and curvature of the metastable energy well smooth the onset of the phase transition. For heterogeneity in $\beta$, increasing variance leads to a complete elimination of the limit-load drop at $\mathrm{CV} = 10.0\%$, resulting in a monotonic response and a stable transition plateau.

In contrast, variability in $\kappa$ and $\mu$ primarily shifts the location of the limit-load drop to earlier stages of compression as $\mathrm{CV}$ increases. This behavior reflects their dominant role in controlling volumetric and shear stiffness rather than directly shaping the non-convex energy landscape. When all parameters are rendered heterogeneous simultaneously, these effects combine synergistically. As the coefficient of variation increases from $0.01\%$ to $1.0\%$, the limit-load drop is reduced and occurs earlier, while at $\mathrm{CV} = 10.0\%$ it disappears entirely. The resulting force--displacement response is smooth and stable, similar to the case of controlling just $\beta$. 

\begin{figure}[htbp]
    \centering
    \includegraphics[width=\linewidth]{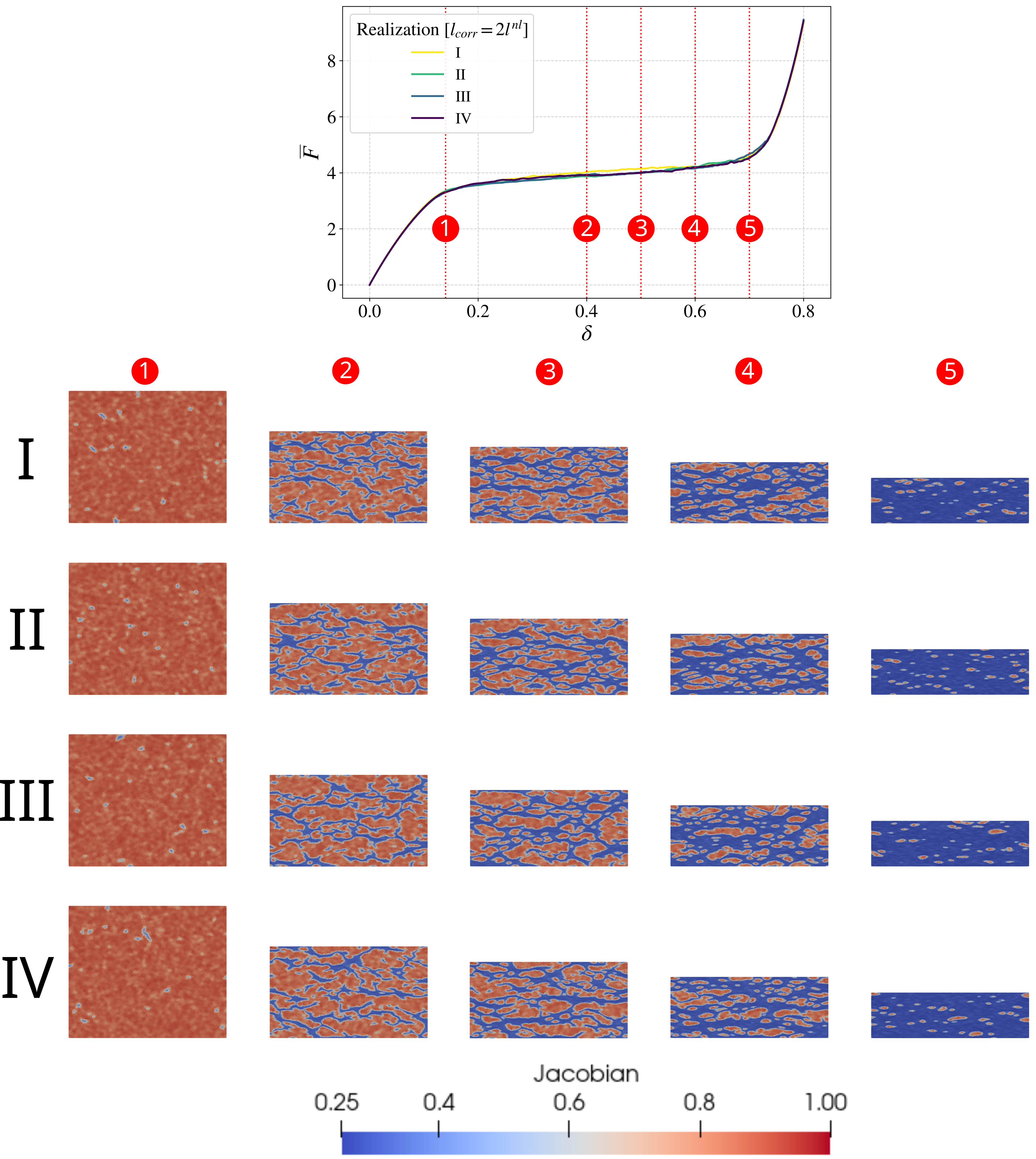}
    \caption{Effect of stochastic heterogeneity on densification patterns under confined compression. Jacobian contours $J$ are shown for four realizations in which all constitutive parameters ($\alpha$, $\beta$, $\kappa$, and $\mu$) are modeled as Gaussian Random Fields with $\mathrm{CV} = 10.0\%$ and correlation length $\ell^{\mathrm{corr}} = 2\ell^{\mathrm{nl}}$ on an unstructured mesh. Five snapshots corresponding to increasing stages of loading illustrating the evolution of the densified phase. Despite realization-to-realization variability in local stiffness distributions, all realizations exhibit consistent macroscopic responses and similar qualitative growth mechanisms of the densified phases.}
    \label{fig:het_triplets_confined_comp}
\end{figure}

To elucidate the underlying deformation mechanisms, Fig. \ref{fig:het_triplets_confined_comp} presents Jacobian contours for four independent stochastic realizations in which all constitutive parameters are modeled as GRFs with identical statistical descriptors. Despite realization-to-realization variability in the local stiffness distribution, the macroscopic force--displacement responses remain in close agreement, indicating that the overall response is statistically robust. The Jacobian fields reveal that densification preferentially initiates within locally softer regions and subsequently propagates through the specimen as loading increases. Unlike the deterministic gradient case, no single preferred nucleation site (e.g., near the top or bottom) is observed. Instead, densified regions emerge diffusely through the domain, creating an interconnected network of regions where volumetric resistance is progressively lost. This is indicative of the spatially distributed nature of the imposed heterogeneity. Importantly, when compared to the homogeneous material response ($0.0\%$ variation) presented earlier in Fig. \ref{fig:kappa_var_gradient}, the characteristic limit-load drop is no longer observed. The introduction of sufficient stochastic heterogeneity distributes the phase transition process spatially, thereby suppressing abrupt localization events that instantaneously create end-to-end collapse band, and yielding a stable macroscopic response. It is noted that an excessive CV was not required to significantly alter the response characteristics in the macroscopic and microscopic level.

\begin{figure}[h!]
    \centering
    \includegraphics[width=\linewidth]{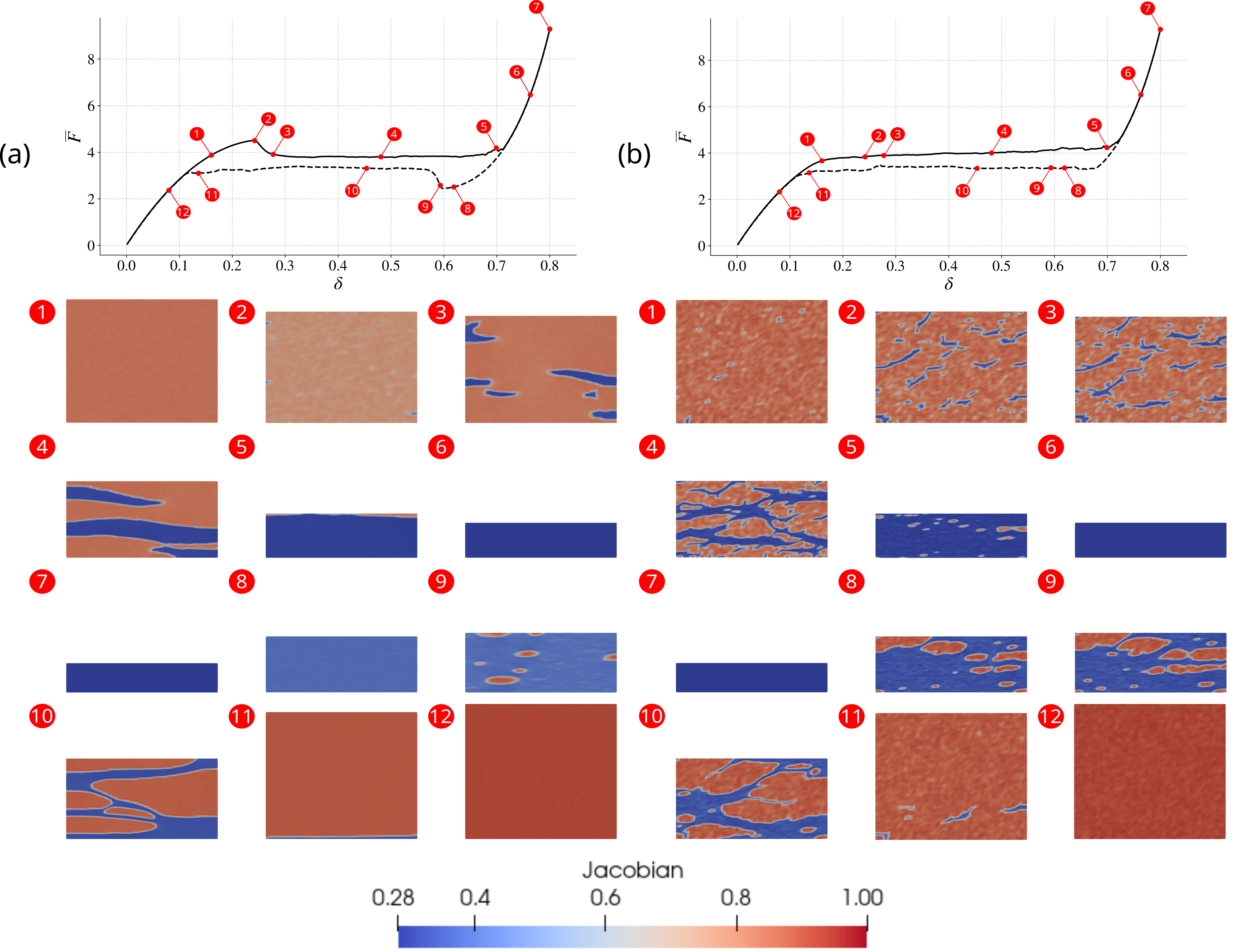}
    \caption{Cyclic confined compression response with stochastic heterogeneity in all constitutive parameters for $\ell^{\mathrm{corr}} = \ell^{\mathrm{nl}}$. Loading--unloading force--displacement curves are shown for (a) $\mathrm{CV} = 1.0\%$ and (b) $\mathrm{CV} = 10.0\%$, together with representative Jacobian contours at selected stages of the cycle. Increasing variance eliminates the limit load drop and leads to a stable hysteresis response, accompanied by a transition from localized band-like densification to distributed, network-like transition patterns.}
    \label{fig:LnU_Contours}
\end{figure}


Finally, Fig. \ref{fig:LnU_Contours} illustrates the cyclic confined compression response when all constitutive parameters are heterogeneous. For $\mathrm{CV} = 1.0\%$, densification patterns resemble those observed under idealized baselines previously studied, with multiple specimen-spanning collapse band indicative of the limit-load observed in the response. These localized bands interact as loading progresses, leading to intermittent force fluctuations. In contrast, for $\mathrm{CV} = 10.0\%$, densification initiates within spatially distributed nuclei of collapsed material, that progressively grow and coalesce to form an interconnected network. No dominant band forms, and the transition proceeds through a gradual growth and coalescence of distributed densified zones. This distributed mechanism eliminates the limit-load drop and produces a stable macroscopic response during both loading and unloading.

\subsection{Effect of lengthscale ratio $\ell^{\text{corr}}/\ell^{\text{nl}}$}\label{Subsection:CorrLength_Var}
As previously discussed, exploring the ratio of the two relevant lengthscales in this problem, namely, the correlation length $\ell^{\text{corr}}$ and the nonlocal lengthscale $\ell^{\text{nl}}$, enables exploring a transition from approximating randomness to stronger correlated structures\footnote{A reminder that the gradient theory introduced here introduces an intrinsic material lengthscale, which through gradient penalization acts like a low-pass filter.}. As such, in this section the influence of the ratio of these lengthscales is examined. In all cases explored here, heterogeneity is introduced in all constitutive parameters using GRFs with a fixed coefficient of variation $\mathrm{CV} = 10.0\%$, while the correlation length is varied as $\ell^{\mathrm{corr}} = \ell^{\mathrm{nl}}$, $2\ell^{\mathrm{nl}}$, $4\ell^{\mathrm{nl}}$, $8\ell^{\mathrm{nl}}$, and $16\ell^{\mathrm{nl}}$. The specimen characteristic dimension $H$ is chosen such that $\ell^{\mathrm{nl}} \ll H$, and the mesh size $h^e$ is chosen such that $h^e=\ell^{\mathrm{nl}}$. The correlation length is chosen in the range $\ell^{\mathrm{nl}} \le \ell^{\mathrm{corr}} < H$. A structured mesh is employed, and for each correlation length four independent stochastic realizations are considered for each numerical experiment. 

\begin{figure}[htbp]
    \centering
    \includegraphics[width=0.8\linewidth]{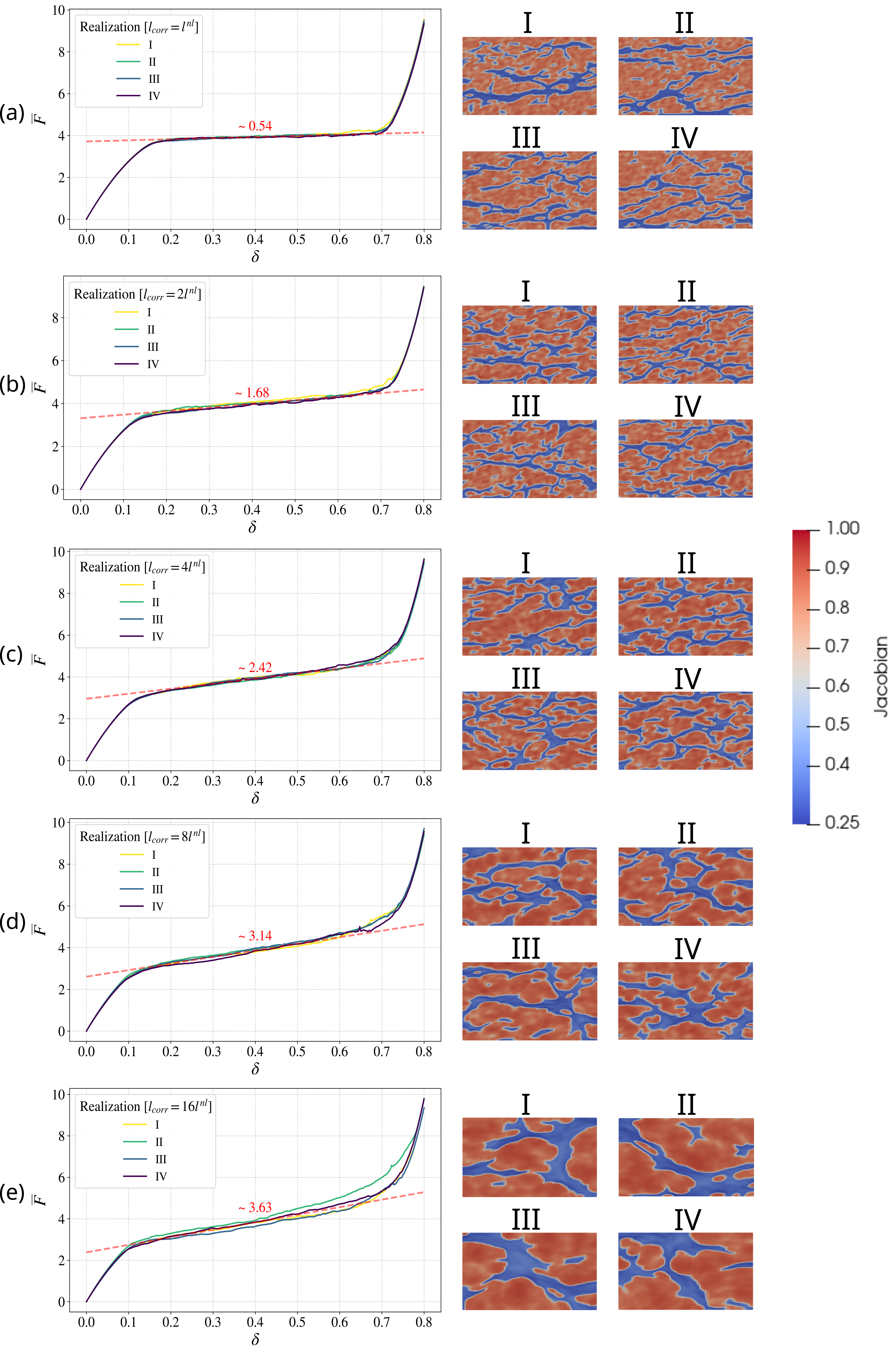}
    \caption{Effect of correlation length on localization under confined compression for $\mathrm{CV} = 10.0\%$ applied to all constitutive parameters. Shown are Jacobian contours at $\delta = 0.4$ for four realizations on a structured mesh with $\ell^{\mathrm{corr}} =$ (a) $\ell^{\mathrm{nl}}$, (b) $2\ell^{\mathrm{nl}}$, (c) $4\ell^{\mathrm{nl}}$, (d) $8\ell^{\mathrm{nl}}$, and (e) $16\ell^{\mathrm{nl}}$. Increasing correlation length leads to larger, more coherent pockets of compliant material and correspondingly coarser densification patterns, resulting in a systematic increase in the intermediate-regime stiffness of the macroscopic response.}
    \label{fig:corr_length_variation_contours}
\end{figure}

Fig. \ref{fig:corr_length_variation_contours} reports snapshots of the Jacobian field at $\delta = 0.4$ for all realizations and correlation lengths. For the shortest correlation length, $\ell^{\mathrm{corr}} = \ell^{\mathrm{nl}}$, the heterogeneity manifests as fine-scale fluctuations, resulting in a highly fragmented densification pattern. As the correlation length is increased to $2\ell^{\mathrm{nl}}$, these pockets grow in size and begin to merge, producing more spatially coherent regions of densification. For $\ell^{\mathrm{corr}} = 4\ell^{\mathrm{nl}}$, the heterogeneity is characterized by large, smoothly varying domains, and the densified phase develops over extended regions rather than through numerous isolated nuclei. These trends are consistent across all realizations, confirming that the observed behavior is governed by the imposed correlation length rather than realization-specific features. 

The influence of correlation length is also reflected in the macroscopic force--displacement response. For $\ell^{\mathrm{corr}} = \ell^{\mathrm{nl}}$, the transition plateau is nearly flat, indicating that the densification proceeds through the progressive activation of many small, weakly interacting regions. For this reason, we quantify the effective tangent stiffness taken as the observed slope during the plateau region in the force displacement plots as seen in \ref{fig:corr_length_variation_contours}. As the correlation length increases to $2\ell^{\mathrm{corr}}$ and $4\ell^{\mathrm{corr}}$, the effective tangent stiffness increases and the macroscopic response is stabilized.  This transition reflects a shift towards collapse band formation that is not spatially uniform, and enables increasing load-carrying capacity throughout the plateau e.g. a uniformly formed densified network that spans the specimen, vs. a densified bands that no longer form in regions perpendicular to the loading direction but follow correlated structures of heterogeneity.

Overall, these results demonstrate that, at a moderately high fixed heterogeneity amplitude (defined through the CV at $10\%$), the correlation length of the underlying GRF plays a central role in controlling both the morphology of localization and the smoothness of the macroscopic response. Large ratios of $\ell^{\text{corr}}/\ell^{\text{nl}}$ promote spatial coherence in the material properties and microscopic deformation patterns, leading to larger densified domains and a more stable, gradually evolving transition plateau under confined compression. This is especially prominent when the correlation length approaches the characteristic size of the domain $\ell^{\text{corr}}\rightarrow H$. Whereas, as small values of $\ell^{\text{corr}}/\ell^{\text{nl}}$ are probed, and a random microarchitecture is approximated, the specimen forms more diffuse collapse band networks and has an almost flat plateau. This trend is expected to hold at lower levels of CV, where a limit load might still be present. 

The work of \cite{hooshmand2022mechanically}, utilized discrete modeling of microarchitecture, highlighted some of the questions that have hopefully been addressed here. Revealing the mechanistic underpinnings of the response is more challenging to probe experimentally, as precise control of the microarchitecture is hard to maintain over large specimens with many unit cells, especially at the level where a continuum is approximated. To further elucidate these trends, Appendix \ref{appendix:Extreme_Heterogeneity} examines large variations in the heterogeneity amplitude and correlation length, highlighting their impact on the macroscopic response. So far, the influence of heterogeneity is showcased towards transitioning from unstable to stable responses, and increasing the slope of the plateau regime. Interestingly, in Appendix \ref{appendix:Extreme_Heterogeneity} the influence of heterogeneity on the extent of the initial elastic regime is also highlighted.

\subsection{Indentation Segment}\label{Sec:Indent_TMC}

We finally examine indentation of the heterogeneous architected metamaterials using a third-medium contact formulation. Contact between the rigid indenter and the specimen is enforced through an auxiliary compliant medium following the approach of Wriggers et al. \cite{wriggers_first_2025}, which provides a smooth and robust treatment of contact constraints within the finite element framework. The governing equations, numerical implementation, and parameter selection for the third-medium formulation are detailed in Appendix \ref{appendix: AM_TMC_Indent}. Here, we focus on how stochastic material heterogeneity influences the indentation response and the associated densification mechanisms as experimental studies have also previously focused on these effects \cite{liang2017compression}, but lacked the resolution to highlight the details that are available in a computational model.
\begin{figure}[h!]
    \centering
    \includegraphics[width=1\linewidth]{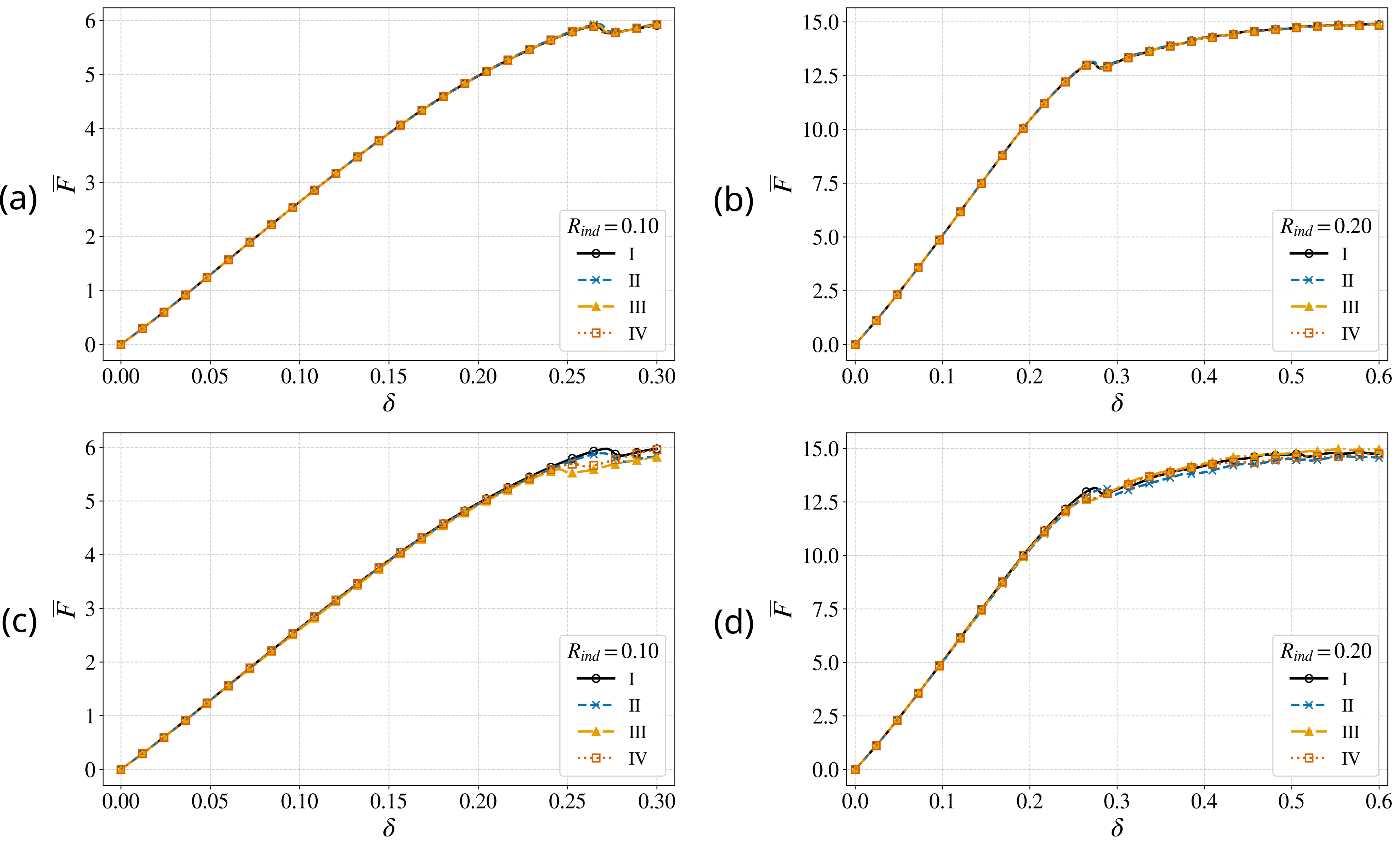}
    \caption{Effect of heterogeneity magnitude and indenter radius on the indentation response. Force--displacement curves are shown for $\mathrm{CV} = 1.0\%$ with (a) $R_{\mathrm{ind}} = 0.1$ and (b) $R_{\mathrm{ind}} = 0.2$, and for $\mathrm{CV} = 10.0\%$ with (c) $R_{\mathrm{ind}} = 0.1$ and (d) $R_{\mathrm{ind}} = 0.2$. In all cases, the correlation length is fixed at $\ell^{\mathrm{corr}} = \ell^{\mathrm{nl}}$, and four independent realizations are reported for each configuration.}
    \label{fig:Indent_CV_and_R_variation}
\end{figure}
All indentation simulations are performed up to a prescribed penetration depth equal to three times the indenter radius, ensuring comparable levels of geometric confinement across all configurations. Fig. \ref{fig:Indent_CV_and_R_variation} summarizes the resulting force--displacement responses for different levels of heterogeneity and indenter radii. For weak heterogeneity ($\mathrm{CV} = 1.0\%$), the force--displacement curves corresponding to different realizations are in near-complete agreement over the entire loading history, indicating that small stochastic fluctuations do not appreciably perturb the macroscopic response. In this regime, densification initiates in a manner closely resembling deterministic indentation-induced localization beneath the indenter. 

As the coefficient of variation is increased to $\mathrm{CV} = 10.0\%$, the responses remain essentially indistinguishable during the initial elastic regime. Differences across realizations emerge only after densification begins to develop, reflecting the sensitivity of the transition process to the spatial arrangement of locally compliant regions introduced by the GRFs. This realization-dependent spread is modest and does not manifest as abrupt force drops, highlighting the stabilizing influence of distributed heterogeneity on the macroscopic response. The effect is more pronounced for the larger indenter radius, $R_{\mathrm{ind}} = 0.2$, where a greater volume of material is engaged and multiple heterogeneous regions are activated simultaneously, increasing competition among densification sites. 

\begin{figure}[h!]
    \centering
    \includegraphics[width=0.7\linewidth]{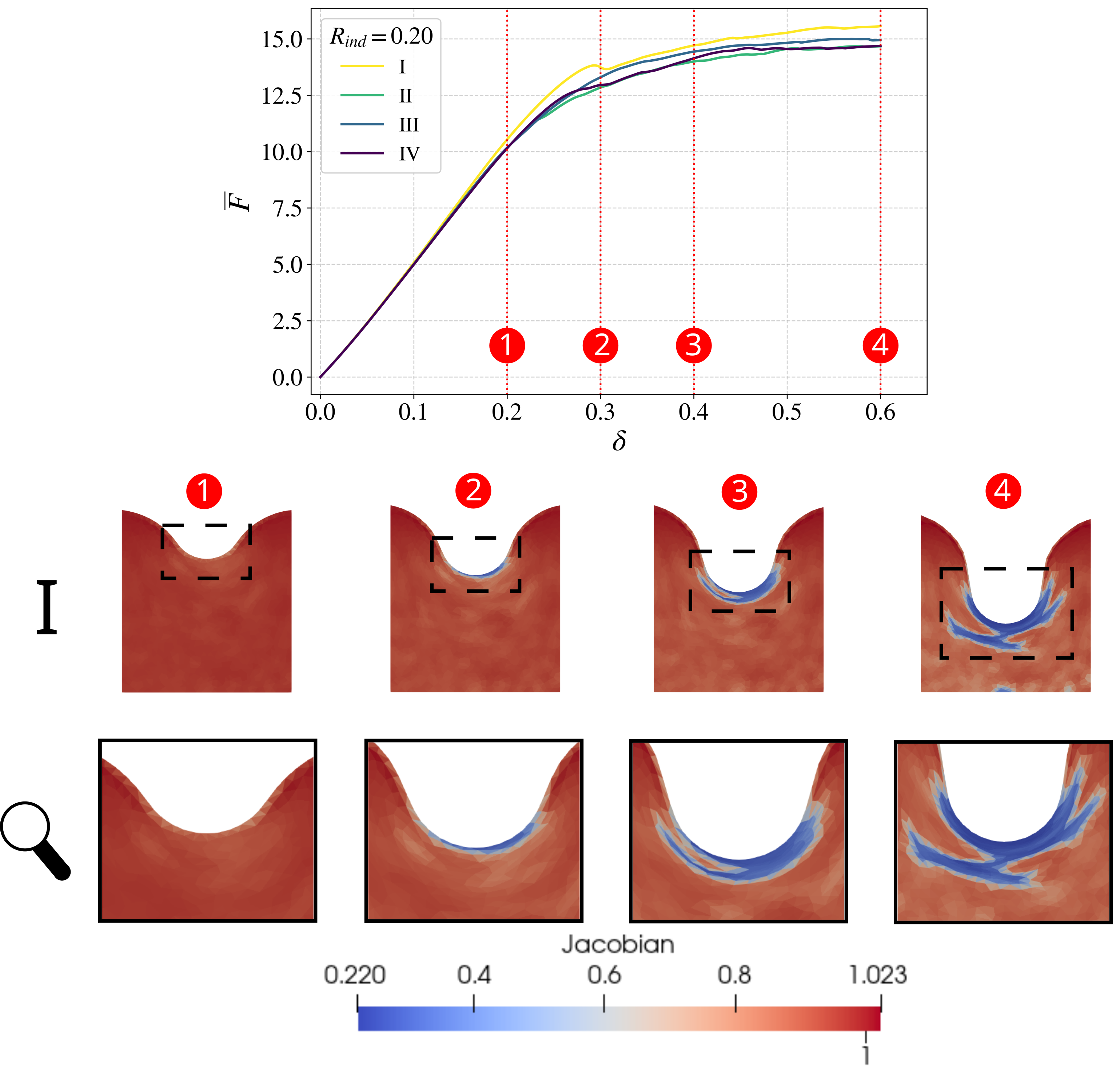}
    \caption{Jacobian contours for indentation with strong heterogeneity. Shown are the four realizations with $\mathrm{CV} = 10.0\%$, correlation length $\ell^{\mathrm{corr}} = 4\ell^{\mathrm{nl}}$, and the indenter radius $R_{\mathrm{ind}} = 0.2$. Snapshots at selected indentation depths illustrate the evolution of densification beneath the indenter, highlighting the formation of spatially distributed, network-like densified regions rather than a single localized band.}
    \label{fig:Indent_Contours_High_R}
\end{figure}

The corresponding evolution of Jacobian field for the strongly heterogeneous, large-indenter case is shown in Fig. \ref{fig:Indent_Contours_High_R}. Although four independent realizations are in the force--displacement response, the contour snapshots correspond to realization I and are representative of the observed behavior. The first row presents global views of the Jacobian field at selected indentation depths, while the second row provides zoomed-in views of the regions directly beneath the indenter. Densification consistently initiates immediately below the indenter, where stresses are highest, a hallmark of indentation-driven response in architected metamaterials \cite{liang2017compression, flores2010indentation, do2020combined}. Owing to the strong heterogeneity and extended correlation length, the densified phase does not condense into a single sharply defined band. Instead, multiple densified regions nucleate within locally compliant domains and evolve into a spatially distributed, network-like morphology.

As indentation progresses, these densified pockets grow and interact, leading to realization-dependent evolution of the densification zone and a corresponding, albeit limited, divergence in the macroscopic response. Importantly, the transition proceeds in a gradual and spatially distributed manner, underscoring the role of strong, correlated heterogeneity in stabilizing indentation-induced phase transitions. 

\begin{figure}[h!]
    \centering
    \includegraphics[width=0.7\linewidth]{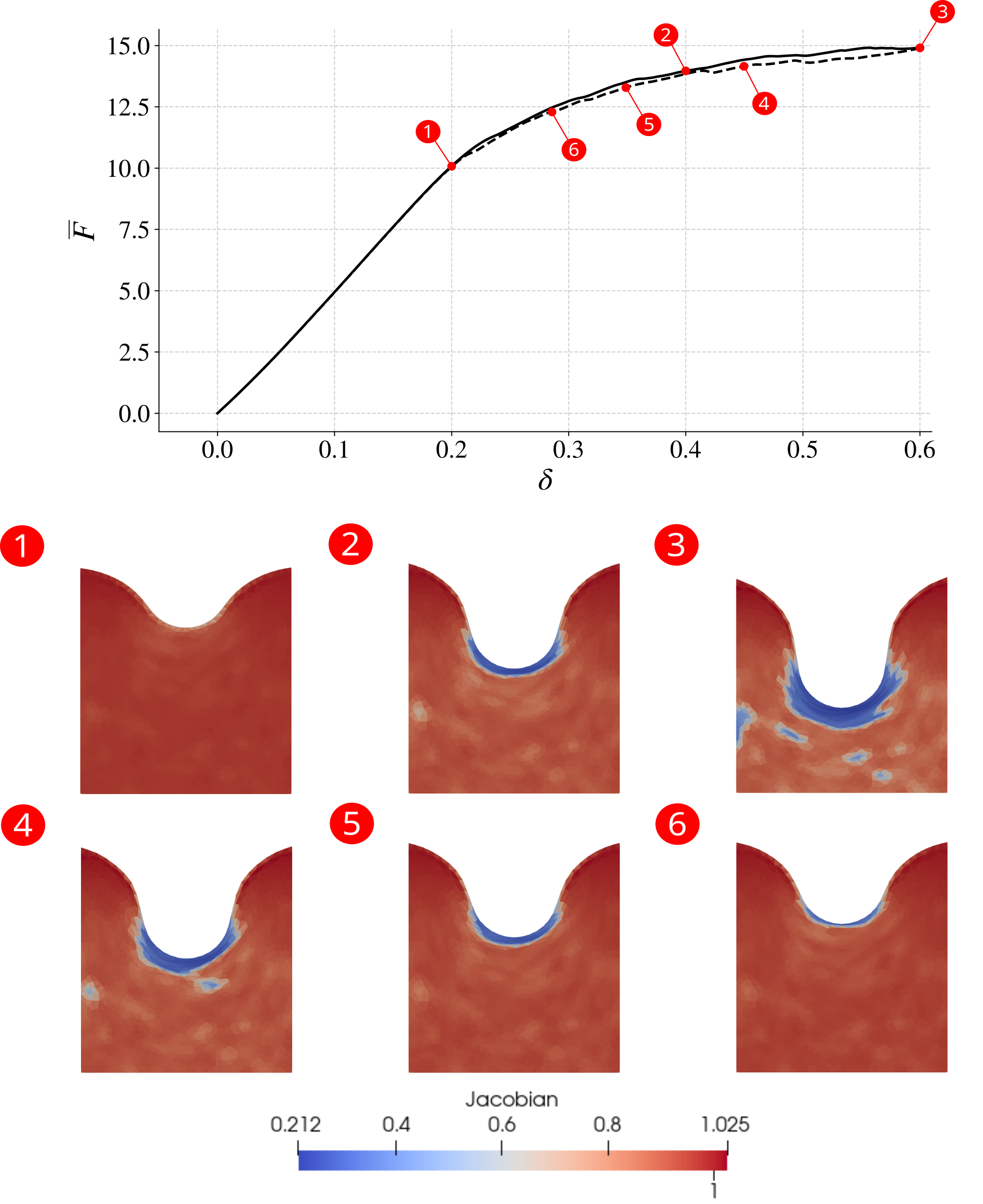}
    \caption{Cyclic indentation response for heterogeneous materials. Loading--unloading force--displacement curves and corresponding Jacobian contours are shown for $\mathrm{CV}=10.0\%$, $\ell^{\mathrm{corr}}=4\ell^{\mathrm{nl}}$, and $R_{\mathrm{ind}}=0.20$. The snapshots highlight the reversible and irreversible components of densification during the indentation cycle.}
    \label{fig:LnU_Indent_Contours}
\end{figure}

Fig. \ref{fig:LnU_Indent_Contours} illustrates the cyclic indentation response for the same configuration. During loading, densification develops progressively beneath the indenter following the distributed patterns described above. Upon unloading, the specimen returns to its original configuration and the macroscopic deformation is fully recovered. The unloading branch nevertheless exhibits hysteresis, consistent with a path-dependent evolution of the densification field during the cycle. The absence of abrupt force drops during both loading and unloading demonstrates that the combination of strong heterogeneity, large correlation length, and finite indenter size promotes a smooth and stable cyclic response. Compared to confined compression, indentation introduces an additional geometric constraint that concentrates deformation near the contact region; nevertheless, stochastic heterogeneity prevents collapse into a single dominant densified region and maintains a distributed morphology. 

Overall, these results demonstrate that, under indentation, stochastic heterogeneity governs both the morphology of densification and the stability of the macroscopic response. Increasing the coefficient of variation, correlation length, and indenter radius promotes smooth indentation behavior by activating densification over extended regions of the material rather than through abrupt localized events.

\section{Conclusion}\label{Section:Conclusion}

This work investigated how stochastic material heterogeneity governs spatial patterning of densification and macroscopic stability in architected metamaterials. The combined effects of heterogeneity amplitude and spatial correlation are studied. The ratio of the governing lengthscales of the system is shown to guide a transition from random microstructures, to strongly correlated microstructures. Using a gradient-enhanced, nonlocal continuum formulation with a non-convex volumetric energy, material disorder was introduced via Gaussian Random Fields characterized by both their coefficient of variation (CV) and correlation length. The results demonstrate that the mechanical response is not controlled by either parameter in isolation, but rather by their coupled interaction with the intrinsic nonlocal regularization length. Highly idealized cases of perfectly homogeneous specimens and specimens with directional gradation of material properties, can show an unstable response. Increasing CV promotes earlier and more spatially distributed phase nucleation and a transition to responses that do not exhibit a limit load and an unstable drop. At the same time, correlation length controls the spatial coherence of these nucleation events and the resulting densification morphology. For small correlation lengths relative to the nonlocal lengthscale, increasing CV leads to highly distributed activation and a nearly flat, stable transition plateau. As the correlation length increases, the same level of heterogeneity produces progressively more coherent collapse patterns and a measurable increase in plateau slope. These findings reveal that macroscopic stability, transition smoothness, and localization topology emerge from a three-way competition between heterogeneity amplitude, heterogeneity correlation length, and nonlocal regularization scale.

Beyond characterizing these trends, the present study resolves several open questions regarding imperfection sensitivity and disorder-driven transitions in architected metamaterials. In particular, it shows that capturing experimentally observed responses requires the simultaneous presence of (i) non-convex energetic landscapes to enable phase transition-like behavior and (ii) spatially heterogeneous material properties to activate and distribute these transitions. Nonlocal regularization alone eliminates mesh dependence and classical imperfection sensitivity but cannot prevent dominant localization in homogeneous non-convex systems. Likewise, introducing heterogeneity within convex or purely local models do not reproduce localized densification and phase-transition-like phenomena. Only when non-convexity, nonlocal regularization, and stochastic heterogeneity-with both finite amplitude and finite correlation length-are combined does the model recover distributed densification, smooth macroscopic transitions, and statistically robust global response, despite strongly heterogeneous local deformation, consistent with experimental observations in foams and architected metamaterials. In this sense, CV and correlation length act as complementary tuning parameters that regulate when transition nucleates and how they spatially organize. In this regard, the limitation of the study to isotropy has to be noted, as well as the fact that the aim was not to recapitulate a specific response of an existing material system.

By showcasing the influence of heterogeneity towards stability, slope of the plateau region, and elimination of the initial elastic regime, compared to that of a homogeneous material that approaches an idealized response, an overarching design paradigm is illuminated in this work. This is enabled by highlighting how heterogeneity can alter features of the macroscopic response. Using heterogeneity to achieve (or understand) target response characteristics is not the prevalent pathway to design of architected metamaterials, as the focus is mostly on unit cell microarchitecture characteristics. 

The framework developed here provides a foundation for several extensions. While the present study focuses on hyperelastic responses, incorporating viscoelastic, viscoplastic, and damage mechanisms would enable modeling of rate effects, irreversible dissipation, and progressive degradation typical of polymeric and metallic foams. Similarly, extensions to anisotropy (in line with the original development of the theory in \cite{joshi2026instabilities} can enable calibration to specific material systems. Materials of interest extend beyond structural and engineered foams to biological materials such as biopolymer networks that are also known to exhibit such phenomena \cite{mollenkopf2025poroelasticity}. 
Finally, data-driven and machine-learning-assisted constitutive modeling offers a promising route for learning effective non-convex energy landscapes and heterogeneity statistics directly from experiments or high-fidelity simulations. Embedding such learned models within a gradient-enhanced framework would allow predictive modeling across length scales while retaining thermodynamic consistency and physical interpretability.

\section*{Acknowledgments}

S.J. and N.B. were supported through NSF-SNSF collaborative program by the U.S. National Science Foundation under Award No. 2516642.

Sandia National Laboratories is a multimission laboratory managed and operated by National Technology \& Engineering Solutions of Sandia, LLC, a wholly owned subsidiary of Honeywell International Inc., for the U.S. Department of Energy’s National Nuclear Security Administration under contract DE-NA0003525. This paper describes objective technical results and analysis. Any subjective views or opinions that might be expressed in the paper do not necessarily represent the views of the U.S. Department of Energy or the United States Government. 

\newpage

\appendix
\gdef\thesection{\Alph{section}}
\makeatletter
\renewcommand\@seccntformat[1]{Appendix \csname the#1\endcsname.\hspace{0.5em}}
\makeatother

\section{Numerical implementation of GRFs} \label{appendix:Numerical_Impl_GRF}

The stochastic partial differential equation introduced in Sec. \ref{Section:GRF_Implementation} is solved numerically using a finite element discretization on the same mesh employed for the mechanical problem. The GRF $G(\mathbf{X})$ is approximated in a continuous Lagrange space of order one, ensuring compatibility with the gradient-enhanced formulation. 

The discretization of Eq. \ref{Eq:GRF_SPDE} leads to a linear system of algebraic equations of the form
\begin{equation}
    \mathbf{A}\mathbf{g} = \mathbf{b},
\end{equation}
where $\mathbf{g}$ denotes the vector of nodal values of the discretized field $G$, $\mathbf{A}$ consists of the discretized Laplacian and zeroth-order operators, supplemented by a Robin-type boundary term to ensure well-posedness and approximate the decay of correlations beyond the computational domain \cite{daon2016mitigating}, and $\mathbf{b}$ represents the discretized white noise forcing. 

The stochastic forcing is constructed by sampling independent standard normal random variables at quadrature points and projecting them consistently into the finite element space through a mass-lumped approximation. This procedure yields a discrete noise vector with unit variance and statistically uncorrelated components. The implementation follows established approaches for SPDE-based sampling of GRFs in PDE-constrained settings, drawing on methodologies developed within the \texttt{hIPPYlib} framework \cite{VillaPetraGhattas21, VillaPetraGhattas18, VillaPetraGhattas16}.

To generate a realization of the GRF, the right-hand side is assembled as
\begin{equation}
    \mathbf{b} = \mathbf{M}^{1/2}\omega,
\end{equation}
where $\omega$ is a vector of independent standard normal random variables and $\mathbf{M}^{1/2}$ denotes the quadrature-based approximation of the square-root of the mass matrix. The resulting linear system is solved using a conjugate gradient method with algebraic multigrid preconditioning, as provided by the \texttt{FEniCS} backend. 

Independent realizations of the GRF are obtained by repeated sampling of the noise vector and solution of the corresponding linear system. The resulting fields are subsequently mapped to spatially varying material parameters using the lognormal transformation described in Sec. \ref{Section:GRF_Implementation}, ensuring positivity while preserving the prescribed mean and variance.


\section{Effective Bulk Modulus: Derivation and Parametric Analysis} \label{appendix:K_eff_Derivation}


This appendix presents the derivation of the effective bulk modulus $K_{\mathrm{eff}}$ within the constitutive framework introduced in Joshi et al. \cite{joshi2026instabilities}. The derivation focuses on the incremental volumetric stiffness associated with homogeneous perturbations about equilibrium, establishing a direct connection between the curvature of the free energy and the effective bulk response. The resulting expression for $K_{\mathrm{eff}}$, evaluated at the relevant equilibrium states, is subsequently used to examine its parametric dependence.

\subsection{Definition of effective bulk modulus}

For a homogeneous deformation characterized by the Jacobian $J = \det \mathbf{F}$, the hydrostatic pressure is defined as,

\begin{equation}
    p = - \frac{\partial \Psi}{\partial J}
\end{equation}

The effective bulk modulus is then defined as the tangent modulus relating an increment in pressure to an infinitesimal volumetric strain,

\begin{equation}\label{Eq:Keff_general}
    K_{\mathrm{eff}} = J \frac{\partial p}{\partial J} = J^2 \frac{\partial^2 \Psi}{\partial J^2},
\end{equation}

evaluated at the equilibrium state of $J$. This definition applies irrespective of whether the equilibrium corresponds to the reference configuration or a densified state.

\subsection{Isochoric contribution under volumetric deformation}

Although the isochoric part of the free energy density is volume-preserving by construction, it contributes to the volumetric stiffness under homogeneous volumetric deformations. For a purely volumetric deformation $\mathbf{F} = \lambda\mathbf{I}$, with $J = \lambda^3$, the isochoric Neo-Hookean energy reduces to

\begin{equation}
    \Psi_{\mathrm{iso}}(J) = \frac{3\mu}{2}\left(J^{-2/3} - 1\right) - \mu \ln J.
\end{equation}

Differentiation yields
\begin{equation}
    \frac{\partial^2 \Psi_{\mathrm{iso}}}{\partial J^2} = \frac{2\mu}{9} J^{-4/3},
\end{equation}

and substitution in Eq. \ref{Eq:Keff_general} gives the isochoric contribution to the effective bulk modulus,

\begin{equation}
    K_{\mathrm{eff}} = \frac{2}{3}\mu J^{-2/3}.
\end{equation}

In particular, at the reference configuration $(J = 1)$,
\begin{equation}
    K_{\mathrm{eff}}^{\mathrm{iso}} = \frac{2\mu}{3}
\end{equation}

\subsection{Volumetric contribution}

The volumetric part of the free energy density consists of a convex logarithmic penalty and a non-convex metastable contribution. The logarithmic term contributes a constant stiffness,

\begin{equation}
    K_{\mathrm{eff}}^{\mathrm{log}} = \kappa,
\end{equation}

independent of the equilibrium state. 

The non-convex volumetric contribution yields,
\begin{equation}
    K_{\mathrm{eff}}^{\mathrm{NC}}(J) = \alpha J^2 \left[\left(\frac{\partial \Psi}{\partial J}\right)^2 + \Phi \frac{\partial^2 \Phi}{\partial J^2}\right],
\end{equation}

where, 
\begin{equation}
    \Phi = \frac{(1 - J)^2}{2} + \beta(1 - J),
\end{equation}

which depends explicitly on the local curvature of the non-convex energy landscape evaluated at the equilibrium value of $J$. At the reference configuration, this contribution reduces to

\begin{equation}
    K_{\mathrm{eff}}^{\mathrm{NC}} = \alpha\beta^2
\end{equation}

\subsection{Reference and dense equilibria}

Combining the above contributions, the effective bulk modulus at the stress-free reference configuration is

\begin{equation}
    K_{\mathrm{eff}}(J = 1) = \frac{2\mu}{3} + \kappa + \alpha\beta^2
\end{equation}

For a densified equilibrium state $(J = J_d < 1)$, the effective bulk modulus follows direcrly from the general definition,

\begin{equation}
    K_{\mathrm{eff}}(J_d) = J_d^2 \frac{\partial^2 \Psi}{\partial J^2}(J_d),
\end{equation}

where, $J_d$ satisfies the equilibrium condition,

\begin{equation}
    \frac{\partial \Psi}{\partial J}(J_d) + p = 0 \qquad \frac{\partial^2 \Psi}{\partial J^2}(J_d) > 0
\end{equation}

In this case, the dependence of $K_{\mathrm{eff}}$ on $\alpha$ and $\beta$ reflects the local curvature of the same non-convex volumetric energy evaluated at a different stationary point, explaining qualitative differences between the reference and dense tangent stiffnesses. 


\subsection{Parametric Study of $\alpha$, $\beta$, $\kappa$ for $\mu = 1.0$}\label{Sec:Parametric_Keff}

The foregoing derivation, culminating in Eq. \ref{Eq:K_eff_Equilibrium}, provides an explicit expression for the effective bulk modulus, $K_{\mathrm{eff}}$, which directly reflects the local curvature of the volumetric free energy evaluated at the corresponding equilibrium state. While this expression characterizes the incremental volumetric stiffness analytically, its implications are best understood through systematic evaluations across the governing parameter space. 

Accordingly, we examine the parametric dependence of $K_{\mathrm{eff}}$ on $\alpha$, $\beta$, and $\kappa$ under purely volumetric loading. In what follows, the shear modulus is fixed to $\mu = 1.0$, and the analysis focuses on how these parameters-controlling the strength, asymmetry, and baseline curvature of the volumetric energy landscape-influence the incremental compressibility.

We first consider the stress-free equilibrium configuration at $J = 1$. Linearization of the Helmholtz free energy density with respect to homogeneous volumetric perturbations yields the effective bulk modulus
\begin{equation}\label{Eq:K_eff_Equilibrium}
    K_{\mathrm{eff}} = \frac{2}{3}\mu + \kappa + \alpha\beta^2\,.
\end{equation}
The term $\frac{2}{3}\mu$ originates from the volumetric coupling of the isochoric Neo-Hookean contribution, $\kappa$ corresponds to the intrinsic logarithmic volumetric penalty, and the term $\alpha\beta^2$ reflects the curvature induced by the non-convex volumetric energy at the reference configuration.
\begin{figure}[h]
    \centering
    \includegraphics[width=\linewidth]{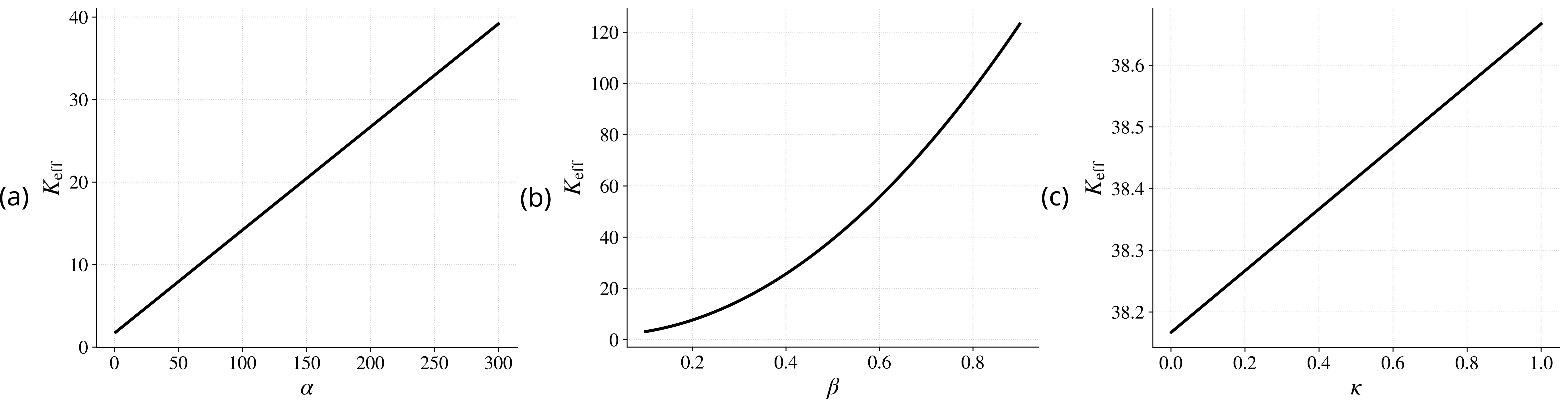}
    \caption{Variation of the effective bulk modulus $K_{\mathrm{eff}}$ with (a) $\alpha$, (b) $\beta$, and (c) $\kappa$, evaluated at the equilibrium state $J = 1$ for $\mu = 1.0$.}
    \label{fig:Equilibrium_K_eff}
\end{figure}
The parametric trends predicted by Eq. \ref{Eq:K_eff_Equilibrium} are summarized in Fig. \ref{fig:Equilibrium_K_eff}. Increasing $\alpha$ leads to a linear increase in $K_{\mathrm{eff}}$. 
In contrast, variations in $\beta$ produce a pronounced nonlinear increase, consistent with its quadratic appearance in Eq. \ref{Eq:K_eff_Equilibrium} and its role in shifting the non-convex landscape relative to the reference state. The dependence on $\kappa$ is strictly linear, as expected from its additive contribution to the volumetric stiffness.

We next evaluate the incremental stiffness about a densified equilibrium state $J = J_d < 1$, defined as a stable stationary point of the reduced volumetric energy under a prescribed compressive hydrostatic loading. In this case, the effective bulk modulus is obtained by linearizing the volumetric response about $J = J_d$, with $J_d$ determined implicitly from the equilibrium condition
\begin{equation}
    \frac{\partial \Psi}{\partial J}(J_d) + p = 0,
\end{equation}
subject to $\partial^2 \Psi /\partial J^2 (J_d) > 0$. The resulting tangent stiffness reflects both the local curvature of the energy landscape and the position of the dense equilibrium. 

\begin{figure}[h]
    \centering
    \includegraphics[width=\linewidth]{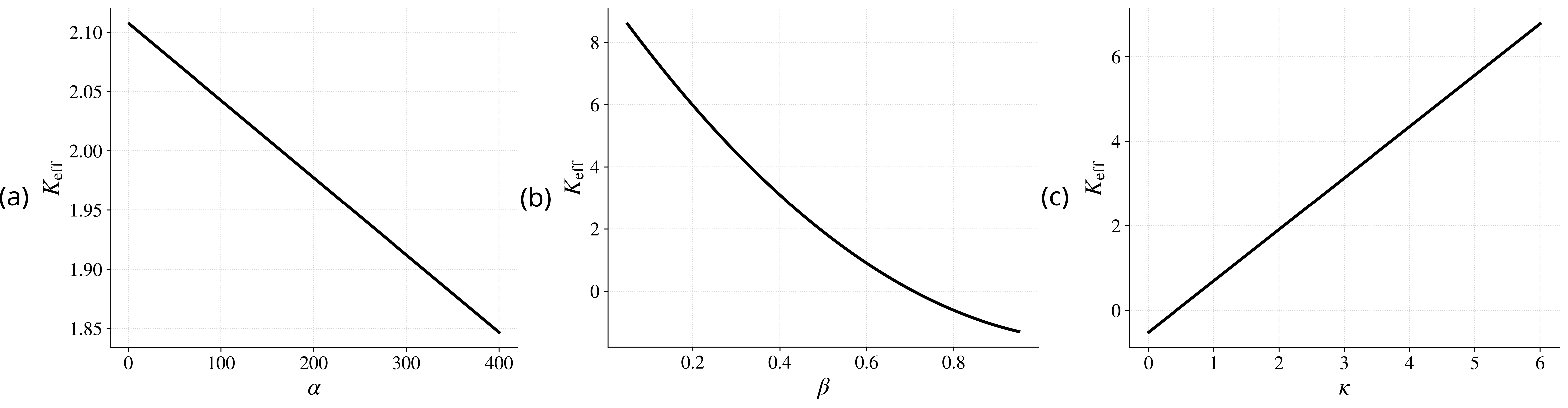}
    \caption{Variation of the effective bulk modulus $K_{\mathrm{eff}}$ with (a) $\alpha$, (b) $\beta$, and (c) $\kappa$, evaluated about the dense equilibrium state $J = 0.2390$ at hydrostatic pressure $p = 5.0$.}
    \label{fig:Dense_K_eff}
\end{figure}

Fig. \ref{fig:Dense_K_eff} shows that the parametric sensitivity of $K_{\mathrm{eff}}$ in the dense phase differs qualitatively from that at the reference state. While increasing $\kappa$ continues to increase the tangent stiffness, increases in $\alpha$ and $\beta$ lead to a reduction of $K_{\mathrm{eff}}$ at $J = 0.2390$. This behavior reflects the fact that $\alpha$ and $\beta$ modify the curvature of the non-convex volumetric energy in a state-dependent manner: while they increase $\partial^2 \Psi/\partial J^2$ at the reference configuration, their influence on $\partial^2 \Psi/\partial J^2$ evaluated at the dense equilibrium $J < 1$ can be opposite, owing to the parameter-dependent shift of the stable minimum.

The origin of these trends are clarified by examining the free energy density $\Psi(J)$, the corresponding Gibbs free energy density $\widehat{\Psi}(J)$, defined as the Legendre transform of $\Psi$ -- corresponding to a transition from displacement-controlled to force-controlled loading, and the associated stress--stretch responses under idealized loading paths. 
\begin{figure}[h]
    \centering
    \includegraphics[width=\linewidth]{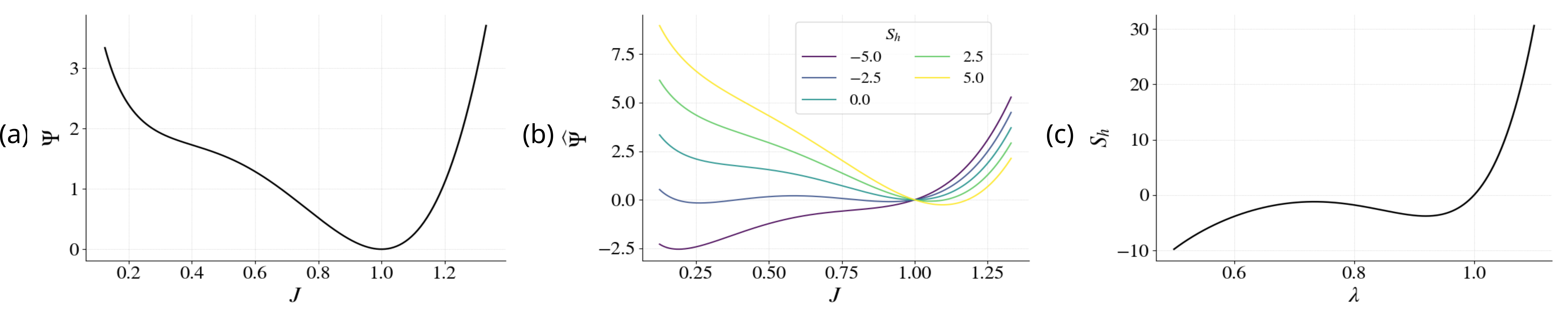}
    \caption{Analytical plots of (a) Helmholtz free energy density $\Psi(J)$, (b) Gibbs free energy density $\widehat{\Psi}(J)$, and (c) hydrostatic stress response $S_h(\lambda)$, defined as the hydrostatic component of the second Piola-Kirchoff stress evaluated under homogeneous hydrostatic deformation $(J = \lambda^3)$.}
    \label{fig:Hydrostatic_Plots}
\end{figure}
Under hydrostatic compression, as illustrated in Fig. \ref{fig:Hydrostatic_Plots}, the non-convex volumetric energy admits both a reference equilibrium near $J = 1$ and a secondary dense basin at $J < 1$. Tilting the energy landscape through the applied hydrostatic stress progressively stabilizes the dense state, leading to a non-monotone stress--stretch response characteristic of metastable volumetric behavior. 
\begin{figure}[h]
    \centering
    \includegraphics[width=\linewidth]{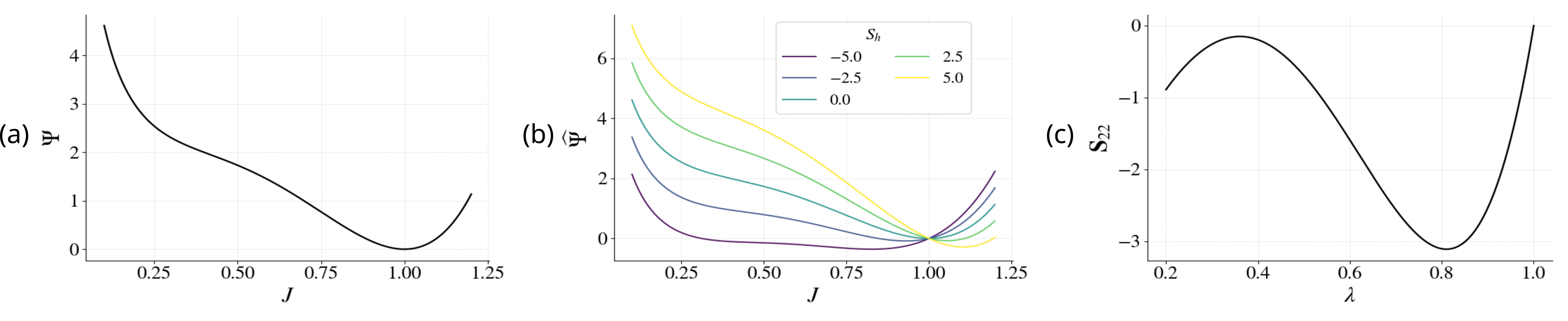}
    \caption{Analytical plots of (a) Helmholtz free energy density $\Psi(J)$, (b) Gibbs free energy density $\widehat{\Psi}(J)$, and (c) referential stress response ($\mathbf{S}_{22}$) under confined compression.}
    \label{fig:Confined_Comp_Plots}
\end{figure}
For confined compression, as illustrated in Fig. \ref{fig:Confined_Comp_Plots}, the kinematic constraint modifies the relation between stretch and volume change, altering how external work biases the energy landscape. As a result, the location of the stable equilibria and the extent of the non-monotone regime differ from the hydrostatic case. These differences underscore that the effective bulk modulus extracted from linearization is inherently state-dependent, reflecting the local curvature of the same non-convex energy landscape evaluated along different loading paths.


\section{A Broader Exploration of Heterogeneity} \label{appendix:Extreme_Heterogeneity}

Building on Sec. \ref{Section:Results}, where we focused on systematic trends associated with moderate variations in the coefficient of variation ($\mathrm{CV}$) and correlation length $\ell^{\mathrm{corr}}$, the present appendix explores representative extreme cases. In particular, we consider highly heterogeneous realizations with large $\mathrm{CV}$, as well as configurations with $\ell^{\mathrm{corr}}$ spanning several orders of magnitude relative to the intrinsic nonlocal length $\ell^{\mathrm{nl}}$. These cases serve to highlight the mechanisms through which heterogeneity amplitude and spatial correlation govern the macroscopic response and the morphology of densification.  

\begin{figure}[h!]
    \centering
    \includegraphics[width=0.8\linewidth]{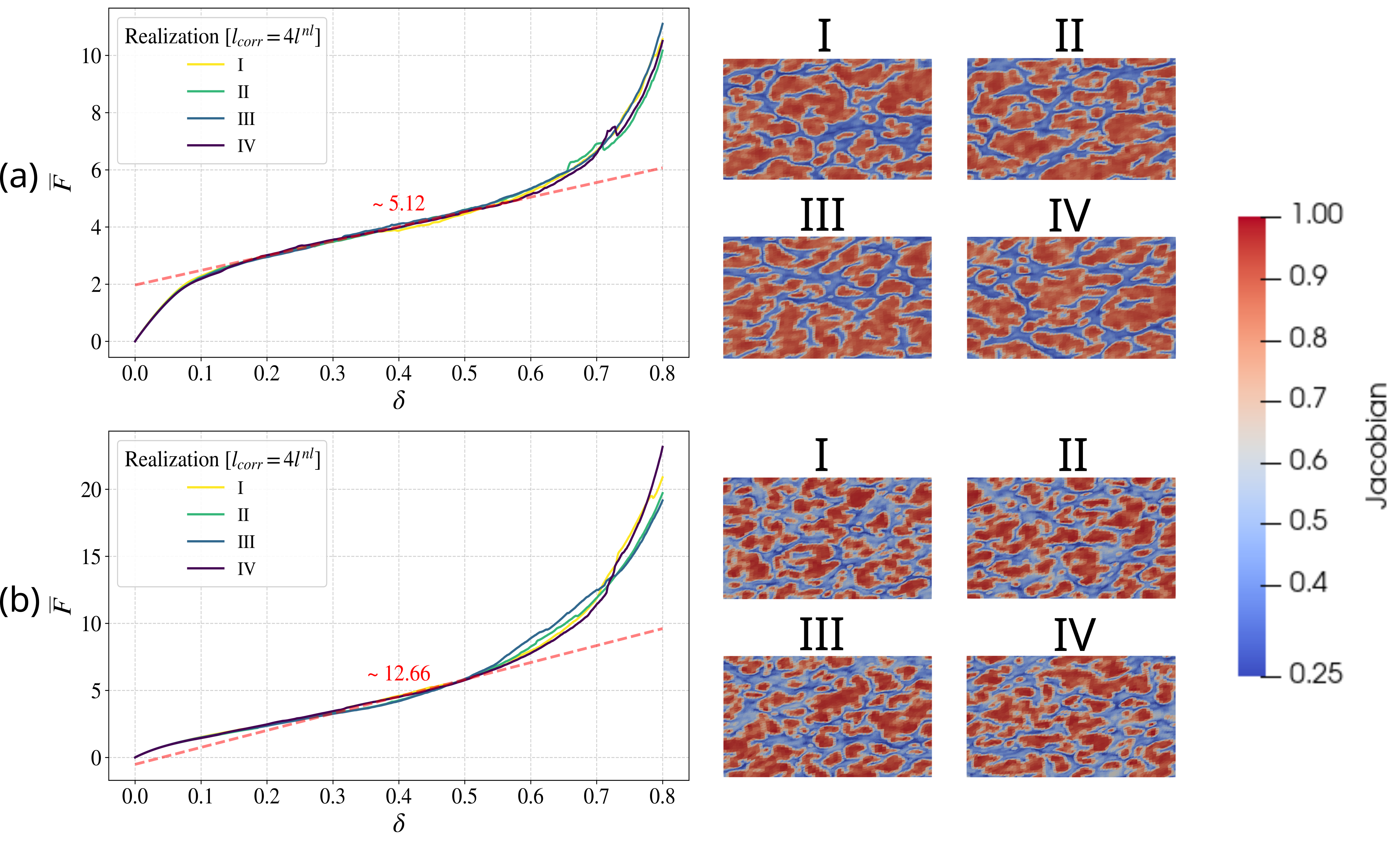}
    \caption{Effect of high heterogeneity amplitude on localization under confined compression for $\ell^{\mathrm{corr}} = 4\ell^{\mathrm{nl}}$ applied to all constitutive parameters. Shown are Jacobian contours at $\delta = 0.4$ for four realizations on a structured mesh with $\mathrm{CV} =$ (a) $20.0\%$ and (b) $50.0\%$. Increasing $\mathrm{CV}$ leads to highly irregular and spatially distributed densification patterns, reflecting the activation of strongly heterogeneous regions and the emergence of asynchronous phase transitions across the specimen.}
    \label{fig:High_CV_Contours}
\end{figure}

We first consider the effect of large heterogeneity amplitude, as illustrated in Fig. \ref{fig:High_CV_Contours}, where $\mathrm{CV}$ values of $20.0\%$ and $50.0\%$ are examined. As the $\mathrm{CV}$ increases, a pronounced reduction in the initial stiffness is observed, reflecting the presence of highly compliant regions that deform at lower loads and weaken the effective elastic response. It is also noted that, for these simulations, the correlation length is fixed at $\ell^{\mathrm{corr}} = 4\ell^{\mathrm{nl}}$, leading to a spatially coherent organization of the heterogeneous fields. The tilted plateau starts earlier for larger CV, resulting in a reduction of macroscopic strength. After the initial linear elastic regime, and tilted plateau, the force--displacement curves exhibit noticeable fluctuations, including intermittent drops in the densification regime. These features arise from the strongly heterogeneous distribution of local properties, which leads to spatially asynchronous phase transitions: regions that remain in the rare phase persist alongside already densified zones, and subsequently undergo abrupt transitions under continued loading.  

\begin{figure}[h!]
    \centering
    \includegraphics[width=0.8\linewidth]{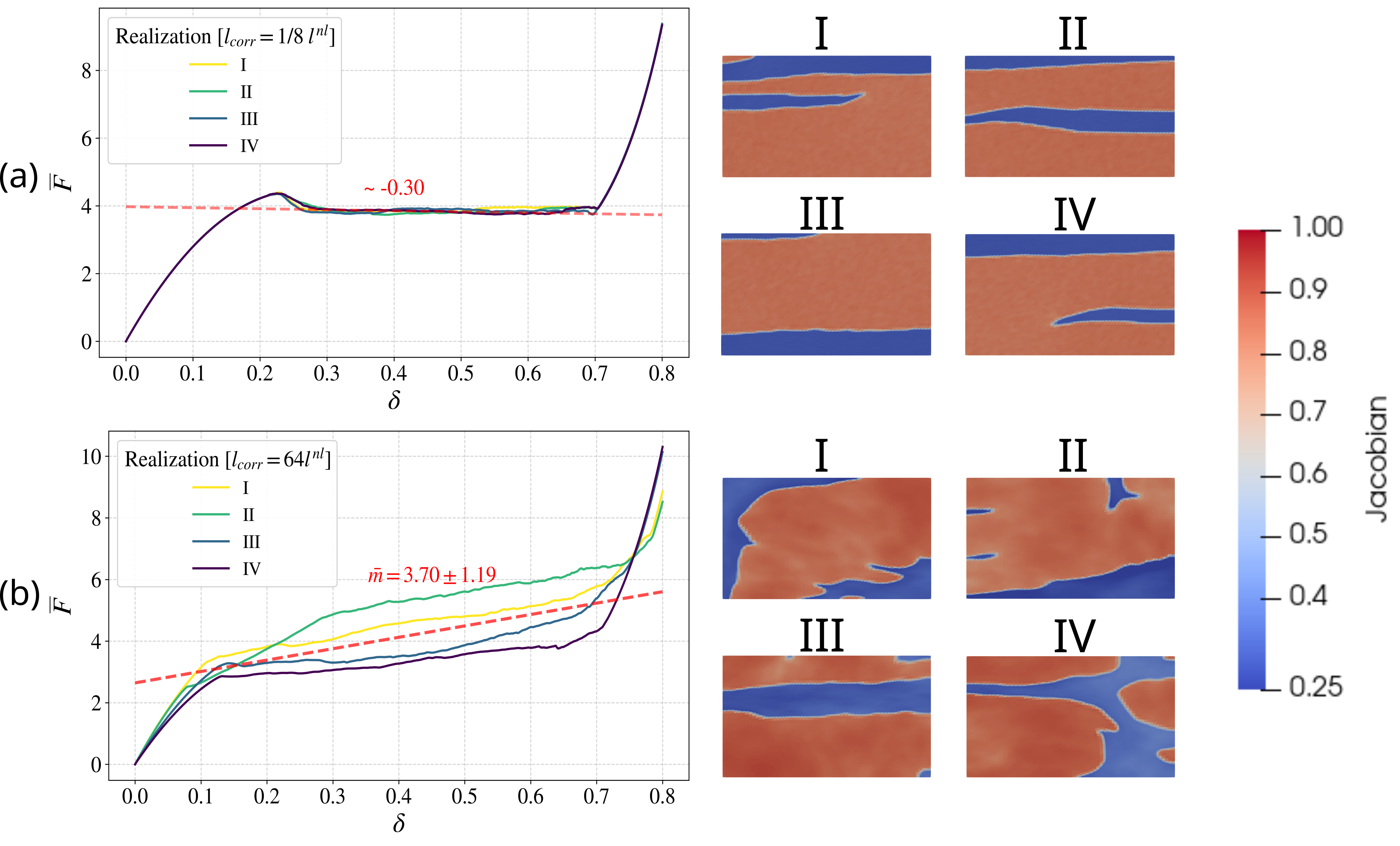}
    \caption{Effect of extreme correlation length $\ell^\mathrm{corr}$ on localization under confined compression for $\mathrm{CV} = 10.0\%$ applied to all constitutive parameters. Shown are Jacobian contours at $\delta = 0.4$ for four realizations on a structured mesh with $\ell^\mathrm{corr} =$ (a) $1/8\;\ell^{\mathrm{nl}}$ and (b) $64\ell^{\mathrm{nl}}$. Decreasing $\ell^{\mathrm{corr}}$ leads to finely distributed heterogeneity and activation of many weakly interacting regions, resulting in statistically consistent macroscopic responses, while increasing $\ell^{\mathrm{corr}}$ produces spatially coherent domains and realization-dependent localization behavior.}
    \label{fig:Extreme_Corr_Lengths}
\end{figure}

We next examine a broader range for the correlation length while fixing the heterogeneity amplitude at $\mathrm{CV} = 10.0\%$. As shown in Fig. \ref{fig:Extreme_Corr_Lengths}, when the correlation length is taken to be $\ell^{\mathrm{corr}} = 1/8\;\ell^{\mathrm{nl}}$, effectively representing a random microstructure, the response exhibits a nearly flat transition plateau, with all realizations showing close agreement, consistent with what was observed for homogeneous or nearly homogeneous specimens. 

In contrast, for $\ell^{\mathrm{corr}} = 64\ell^{\mathrm{nl}}$, where the correlated microstructural zones are at the structural scale, the response becomes significantly more variable across realizations. The force-displacement curves exhibit distinct transition behaviors, including variations in the onset of localized compaction, differences in the extent of the plateau region, and changes in its slope and curvature. In particular, some realizations show an early onset of the transition, while others exhibit delayed densification, leading to a spread in the effective tangent stiffness of the plateau, with a mean value of approximately $3.70$ and a standard deviation of $1.39$. These variations arise from the strong coherence of the heterogeneity at large correlation lengths, where the response is governed by a small number of dominant regions, making the macroscopic behavior sensitive to realization-specific features.

These cases extend the findings from the main text, confirming that the governing roles of $\mathrm{CV}$ and $\ell^{\mathrm{corr}}$ persist even when exploring a broader range of values.


\section{Third-Medium Contact Formulation for Indentation} \label{appendix: AM_TMC_Indent}

This appendix summarizes the variational formulation of the third-medium contact approach used to model indentation of heterogeneous metastable solid in Sec. \ref{Sec:Indent_TMC}. The formulation follows the third-medium concept introduced in Wriggers et al. \cite{wriggers2025third}, in which contact is enforced through an auxiliary deformable continuum occupying the potential contact region. The auxiliary medium acts as an energetic barrier: it contributes negligibly in separation, while developing a rapidly increasing resistance under compression, thereby preventing interpenetration without invoking inequality constraints. 

\subsection{Geometric setting and kinematics}

We consider a two-dimensional plane strain setting. The architected metamaterial occupies the domain
\[
\Omega_s = (0,1)\times(0,1),
\]
while the third-medium occupies
\[
\Omega_m = (0,1)\times\big(1,\,1+2R_{\mathrm{ind}}+0.02\big)\setminus\mathcal{C},
\]
where $\mathcal{C}$ denotes a cavity representing the rigid indenter. The punch boundary is given by $\Gamma_p = \partial\mathcal{C}$. The third-medium thus forms a finite buffer region surrounding the indenter cavity rather than a thin interfacial layer. 

The deformation of the combined domain $\Omega_0 = \Omega_s\cup\Omega_m$ is described by the displacement field $\mathbf{u}$. The deformation gradient is defined as,
\[
\mathbf{F} = \mathbf{I} + \nabla\mathbf{u}, \qquad \text{with}\ J = \det \mathbf{F}, \qquad \mathbf{C} = \mathbf{F}^T\mathbf{F}. 
\]

Under plane strain embedding, the first invariant
\[
I_1 = \mathrm{tr}\mathbf{C} + 1
\]
is employed.

\subsection{Helmholtz free energy density of the solid}
The constitutive response of the architected metamaterial follows the gradient-enhanced metastable formulation introduced in Sec. \ref{Section:Nonlinear_Theory}. In addition to the displacement field $\mathbf{u}$, the solid is endowed with a nonlocal volumetric internal variable $\tilde{J}$, which regularizes volumetric localization and enables metastable phase transitions. 

The Helmholtz free energy density of the solid is given by,
\[
\Psi_s(\mathbf{F}, \tilde{J}, \nabla\tilde{J}) = \frac{\mu}{2}(I_1 - 3 - 2\ln J) + \frac{\kappa}{2}(\ln J)^2 + \frac{\alpha}{2}\left(\frac{(1 - \tilde{J})^2}{2} + \beta(1 - \tilde{J})\right)^2 + c(J - \tilde{J})^2 + d{\ell^{\mathrm{nl}}}^{2}\|\nabla\tilde{J}\|^2
\]
with, $\mu$ and $\kappa$ denote the shear and bulk moduli, $\alpha$ and $\beta$ govern the non-convex volumetric contribution, $c > 0$ penalizes deviations between $J$ and $\tilde{J}$, and $\ell^{\mathrm{nl}}$ is the nonlocal length scale. Artificial viscosity is introduced in the evolution equation for $\tilde{J}$, as described in Sec. \ref{Section:Nonlinear_Theory}, to stabilize the propagation of volumetric phase fronts.

\subsection{Third-medium barrier energy}

Contact is enforced by introducing an auxiliary third-medium occupying $\Omega_m$. The third-medium is designed such that its energetic contributions are negligible in separation, while it becomes increasingly stiff under compression.

The bulk response of the third-medium is described by the isochoric barrier energy
\[
\Psi_m^{\mathrm{iso}}(\mathbf{F}) = \frac{G}{2}\left(J^{-2/3}I_1 - 3\right), \qquad G = \frac{E}{2(1 + \nu)}
\]
where $E$ and $\nu$ are elastic parameters associated with the third-medium. This contribution is scaled by a small factor $\gamma_\mathrm{tm}\ll 1$.

To ensure numerical robustness under large distortions, the barrier energy is augmented by mixed regularization terms involving an auxiliary scalar field $p$. This nomenclature is consistent with Wriggers et al. \cite{wriggers2025third} and is not meant to be interpreted as hydrostatic pressure. Introducing the stabilized shear/rotation measure
\[
\tau(\mathbf{F}) = \frac{F_{12} - F_{21}}{F_{11} + F_{22}},
\]
the regularization energy is defined as
\[
\Psi_m^{\mathrm{reg}}(\mathbf{F}, \tilde{J}, p, \nabla\tilde{J}, \nabla p) = \frac{\beta_1}{2}\left(\tau(\mathbf{F}) - p\right)^2 + \frac{\alpha_r}{2}\|\nabla p\|^2 + \frac{\beta_2}{2}\left(J - \tilde{J}\right)^2 + \frac{\alpha_r}{2}\|\nabla\tilde{J}\|^2.
\]
The coupling between $J$ and $\tilde{J}$ here ensures a consistent transmission of volumetric constraints across the solid-third-medium interface. 

The total potential energy of the coupled system reads
\[
\Pi(\mathbf{u},\tilde{J},p)
=
\int_{\Omega_s}\psi_s\,\mathrm{d}V
+
\gamma_{\mathrm{tm}}\int_{\Omega_m}
\big(\psi_m^{\mathrm{iso}}+\psi_m^{\mathrm{reg}}\big)\,\mathrm{d}V
+
\frac{\varepsilon_{\mathrm{out}}}{2}\int_{\Omega_s}p^2\,\mathrm{d}V.
\]

The final term weakly anchors the auxiliary variable $p$ inside the solid domain, where it has no physical interpretation, and is introduced solely to eliminate null modes in the mixed formulation. The parameter $\varepsilon_{\mathrm{out}}\ll 1$ is chosen sufficiently small so as to not influence the mechanical response. 

\subsection{Reaction force}

The indentation reaction is obtained from the vertical traction transmitted across the internal interface
\[
\Gamma_{\mathrm{int}} = \partial\Omega_s\cap\partial\Omega_m.
\]
Denoting by $\mathbf{P}$, the first Piola-Kirchoff stress associated with $\Psi_s$, the reaction force is defined as
\[
F_y(\delta) = -\int_{\Gamma_{\mathrm{int}}} P_{22}\,\mathrm{d}S.
\]
This definition yields the force--displacement responses reported in Sec. \ref{Sec:Indent_TMC}. 


\newpage
\FloatBarrier

\end{document}